\documentclass[aps,prd,11pt,notitlepage,longbibliography,nofootinbib,tightenlines,preprintnumbers,superscriptaddress]{revtex4-2}
\pdfoutput=1

\makeatletter
\renewcommand{\p@subsection}{}
\renewcommand{\p@subsubsection}{}
\makeatother

\usepackage{amsmath,amssymb,amsfonts}
\usepackage{graphicx}
\usepackage{color}
\usepackage{tikz}
\usepackage{dsfont}
\usepackage[pdftex]{hyperref}
\hypersetup{colorlinks=true, linkcolor=darkred, citecolor=blue, linktoc=page}
\definecolor{darkred}{rgb}{0.8,0.1,0.1}

\definecolor{3dcolor}{rgb}{0.96,0.89,0.76}
\definecolor{4dcolor}{rgb}{0.812,0.851,0.914}

\usepackage{subfigure}

\def\CC{{\mathds{C}}}
\def\RR{{\mathds{R}}}
\def\ZZ{{\mathds{Z}}}

\def\cA{{\cal A}}
\def\cC{{\cal C}}
\def\cD{{\cal D}}
\def\cY{{\cal Y}}
\DeclareMathOperator{\vol}{vol}

\DeclareMathOperator{\tr}{tr}
\DeclareMathOperator{\sech}{sech}

\makeatletter\def\l@subsubsection#1#2{}%
\makeatother

\newcommand{\nocontentsline}[3]{}
\newcommand{\tocless}[2]{\bgroup\let\addcontentsline=\nocontentsline#1{#2}\egroup}

\def\Im{\mathop{\rm Im}}
\def\Re{\mathop{\rm Re}}

\begin{document}

\title{BMN-like sectors in 4d \texorpdfstring{$\mathcal N=4$}{N=4} SYM with boundaries and interfaces}

\author{Andrea Chaney}
\email{chaneya@umich.edu}

\affiliation{
	Department of Physics, University of Michigan, 450
	Church St, Ann Arbor, MI 48109-1040, USA\\[2mm]}

\author{Christoph F.~Uhlemann} 
\email{christoph.uhlemann@vub.be}

\affiliation{Theoretische Natuurkunde, Vrije Universiteit Brussel and The International Solvay
	Institutes, Pleinlaan 2, B-1050 Brussels, Belgium}

\begin{abstract}
4d $\mathcal N\,{=}\,4$ SYM admits half-BPS boundaries, defects and interfaces as well as compactifications to 3d $\mathcal N\,{=}\,4$ SCFTs, realized as intersections of D3, D5 and NS5 branes. 
We explore operators with large R-charge. We identify Penrose limits in the holographic duals with geometry $\rm AdS_4\times S^2\times S^2\times\Sigma$, which describe, for each theory, a spectrum of `seed' operators which each give rise to a BMN-like sector described by a pp-wave.
We relate the number of BMN-like sectors to the flavor symmetry of the CFT.
For a sample of theories including the Janus interface, D3/D5 BCFTs and BCFTs with 3d SCFTs on the boundary, we determine the spectra of nearby operators and discuss the field theory realization of the seed operators. We further identify families of Penrose limits which give rise to more general pp-wave sectors. 
\end{abstract}

\maketitle
\tableofcontents
\parskip 1mm

\section{Introduction}

Boundaries, defects  and interfaces are important ingredients in quantum field theory, as no real-world system is of infinite extent and truly translation invariant.
In this paper we study boundaries, defects and interfaces in the perhaps simplest interacting 4d QFT, 4d $\mathcal N=4$ SYM.

4d $\mathcal N=4$ SYM admits large classes of boundary conditions, defects and interfaces which preserve 3d $\mathcal N=4$ superconformal symmetry \cite{Gaiotto:2008sa,Gaiotto:2008sd,Gaiotto:2008ak}. 
In string theory these are realized by D3-branes terminating on or intersecting D5 and NS5 branes.
This allows for the realization of 4d boundary CFTs, as well as defects or interfaces between 4d $\mathcal N=4$ SYM theories with independent gauge groups.
The resulting theories generically involve genuine 3d CFTs on the boundaries, defects or interfaces. 
They have known holographic duals, constructed in their general form in \cite{DHoker:2007zhm,DHoker:2007hhe,Aharony:2011yc,Assel:2011xz,Assel:2012cj}, which we will use to study them.
The geometry is a warped product of $\rm AdS_4$ and two $S^2$'s over a Riemann surface $\Sigma$. 
The fully backreacted brane construction is encoded in the choice of $\Sigma$ with a pair of harmonic functions, and the $\rm AdS_5\times S^5$ dual for standard  4d $\mathcal N=4$ SYM is included as a special case.
These dualities have been validated through comparisons of
free energies  \cite{Assel:2012cp,Raamsdonk:2020tin,Coccia:2020wtk} and Wilson loops \cite{Assel:2015oxa,Coccia:2021lpp} to field theory results.
They have been employed e.g.\ in studies of massive gravity \cite{Bachas:2017rch,Bachas:2018zmb},
and in black hole studies and double holography \cite{Uhlemann:2021nhu,Demulder:2022aij,Karch:2022rvr,Deddo:2023oxn}. 

An essential characteristic of the field theories is the spectrum of local operators. Holographic duals in principle give access to the spectrum through supergravity and string fluctuations, whose masses and charges translate to the spectrum of operators. For the theories of interest here, however, extracting this information is challenging, due to the complexity of the holographic duals. Even for supergravity spin-2 fluctuations, which decouple from all other fields \cite{Bachas:2011xa} (see also \cite{DeLuca:2023kjj,Lima:2023ggy}), obtaining the full spectrum generally amounts to solving PDEs on $\Sigma$ which are complicated by the presence of 5-brane sources. As a result, little is known about the Kaluza-Klein spectrum. In this work we use Penrose limits as a tool for spectroscopy of large-charge operators.

We identify operators with large charge under the $SO(3)\times SO(3)$ R-symmetry, represented holographically as strings orbiting the spheres in the $\rm AdS_4\times S^2\times S^2\times\Sigma$ duals with large angular momenta.
The background experienced by such strings is described by Penrose limits of the $\rm AdS_4\times S^2\times S^2\times\Sigma$ solutions, which lead to pp-wave solutions describing the geometry near null geodesics orbiting the $\rm S^2$'s.
Following the work by Berenstein, Maldacena and Nastase (BMN) for $\rm AdS_5\times S^5$ \cite{Berenstein:2002jq}, the pp-wave string vacuum can be identified with a field theory operator with the corresponding charges.
For the CFTs with boundaries and defects considered here, the pp-wave vacua correspond to operators localized on the defect, and broadly identifying null geodesics and Penrose limits provides the spectrum of `seed operators' giving rise to pp-wave sectors.
The 3d $\mathcal N=4$ representations comprise various short and long multiplets \cite{Dolan:2008vc,Cordova:2016emh}, and to broadly capture seed operators corresponding to pp-wave vacua we do not constrain the angular momenta by shortening conditions.
The spectrum of seed operators identified via geodesics provides non-trivial information.
The spectrum of string fluctuations on the pp-wave limits provides in addition the scaling dimensions of `nearby' operators obtained by small modifications of the seed operator.

We will derive general results and exemplify them in a sample of concrete field theories, including the Janus interface, D3/D5 BCFTs, BCFTs with 3d CFTs on the boundary, and interfaces hosting 3d CFTs between N=4 SYM theories on half spaces.
The spectrum of seed operators depends qualitatively on the theory, i.e.\ its brane construction in terms of D3, D5 and NS5 branes and correspondingly on the specific $\rm AdS_4\times S^2\times S^2\times\Sigma$ solution.
We generally find a set of single-charge BPS geodesics for each $SO(3)$ factor. Their numbers can be inferred from the brane construction and field theories. 
Remarkably, despite the reduced supersymmetry of the general $\rm AdS_4\times S^2\times S^2\times\Sigma$ solutions, the pp-wave limits for all single-charge BPS geodesics take the same form as the pp-wave limit of $\rm AdS_5\times S^5$ discussed in \cite{Berenstein:2002jq}.
The (super)symmetry thus enhances in the pp-wave limit\footnote{A similar enhancement was found for  $\rm AdS_5\times M_5$ solutions with reduced supersymmetry whose Penrose limits lead to the same maximally supersymmetric pp-wave solution as $\rm AdS_5\times S^5$ in \cite{Itzhaki:2002kh,Gomis:2002km,PandoZayas:2002dso}.}
and the string spectrum can be obtained explicitly \cite{Metsaev:2001bj}.
For the aforementioned sample of field theories we explicitly derive the spectra of nearby operators, which in general differs for different BMN-like sectors within the same theory, and discuss the field theory realization of the seed operators.

We further identify families of BPS geodesics corresponding to operators in unprotected BPS multiplets, whose scaling dimensions can not be inferred as easily in field theory. We identify various `phase transitions' where the spectra of such operators change qualitatively for the aforementioned sample theories. 
Finally, we identify families of non-BPS geodesics pointing to large-charge operators in long multiplets with specific ratios between the R-charges and scaling dimensions.

\subsection{Overview and outline}\label{sec:geod-overview}

We study configurations of D3, D5 and NS5 branes which realize 4d BCFTs, ICFTs and dCFTs, including the Janus ICFT, D3/D5 BCFTs and the examples in figs.~\ref{fig:D3NS5-brane}, \ref{fig:D3D5NS5-brane}, \ref{fig:D52NS52-brane}. 
The field theories will be discussed in sec.~\ref{sec:fieldtheory} with a summary of the results of the holographic analyses.
In the following we summarize the holographic results, to facilitate a transition directly to sec.~\ref{sec:fieldtheory}.

The ambient 4d $\mathcal N=4$ SYM theories are engineered by D3-branes extending along the (0123) directions, and we take the NS5-branes along (012456) and the D5 along (012789). The 3d $\mathcal N=4$ defect superconformal symmetry has an $\rm SO(3)_C\times SO(3)_H$ R-symmetry. In the brane construction $\rm SO(3)_C$ rotates the NS5 directions transverse to the D3-branes, i.e.\ (456), while $\rm SO(3)_H$ rotates the D5 directions transverse to the D3, i.e.\ (789) \cite[sec.~2]{Hanany:1996ie}.

\begin{figure}
	\subfigure[][]{
		\begin{tikzpicture}[xscale=0.8,yscale=0.9]
			\shade [right color=3dcolor!100,left color=3dcolor!100] (-0.3,0)  rectangle (0.3,-2);
			
			\shade [ left color=3dcolor! 100, right color=4dcolor! 100] (0.3-0.01,0)  rectangle (2,-2);
			\shade [ right color=3dcolor! 100, left color=4dcolor! 100] (-0.3+0.01,0)  rectangle (-2,-2);

			\draw[black!60!white,very thick,dashed] (-0.7,-2) arc (-180:-360:0.45);
			\draw[black!60!white,very thick,dashed] (1.1,-2) .. controls (1.1,-1.3) and (0.4,-0.7) .. (0.4,0);

			\draw[green!70!black,very thick] (-1,-2) arc (-180:-360:0.45);
			\draw[green!70!black,very thick] (-1.5,-2) .. controls (-1.5,-1) and (-0.5,-1) .. (-0.5,0);
			\draw[green!70!black,very thick] (0.8,-2) .. controls (0.8,-1.3) and (0.1,-0.7) .. (0.1,0);

			\draw[thick] (-2,0) -- (2,0);
			\draw[thick] (-2,-2) -- (2,-2);
			\draw[dashed] (2,-2) -- +(0,2);
			\draw[dashed] (-2,-2) -- +(0,2);
			
			\node at (-1,-0.5) {$\Sigma$};
			\node at (2.6,-0.65) {\footnotesize $\rm AdS_5$};
			\node at (2.6,-1) {\footnotesize $\times$};
			\node at (2.6,-1.35) {\footnotesize $\rm S^5$};			
			
			\node at (-2.6,-0.65) {\footnotesize $\rm AdS_5$};
			\node at (-2.6,-1) {\footnotesize $\times$};
			\node at (-2.6,-1.35) {\footnotesize $\rm S^5$};
			
			\foreach \i in {-1.55,-0.5,0.7,1.65} \draw[red,fill=red] (\i,0) 	circle (2pt);
			\foreach \i in {-1.5,-0.1,1.4} \draw[blue,fill=blue] (\i,-2) 	circle (2pt);		
			
			\foreach \i in {-1,...,1}{
				\draw[very thick] (1.2*\i+0.1,-0.08) -- (1.2*\i+0.1,0.08) node [anchor=south] {\footnotesize D5};}
			
			\foreach \i in {-1,1}{
				\draw[thick] (0.9*\i-0.1,-1.92) -- (0.9*\i-0.1,-2.08) node [anchor=north] {\footnotesize NS5};
			}
		\end{tikzpicture}
	}\hskip 8mm
	\subfigure[][]{
		\begin{tikzpicture}[xscale=0.8,yscale=0.9]
			\shade [right color=3dcolor!100,left color=3dcolor!100] (-0.3,0)  rectangle (0.3,-2);
			
			\shade [ left color=3dcolor! 100, right color=4dcolor! 100] (0.3-0.01,0)  rectangle (2,-2);
			\shade [ right color=3dcolor! 100, left color=3dcolor! 100] (-0.3+0.01,0)  rectangle (-2,-2);
			
			\draw[thick] (-2,0) -- (2,0);
			\draw[thick] (-2,-2) -- (2,-2);
			\draw[dashed] (2,-2) -- +(0,2);
			\draw[thick] (-2,-2) -- +(0,2);
			
			\node at (-1,-0.5) {$\Sigma$};
			\node at (2.6,-0.65) {\footnotesize $\rm AdS_5$};
			\node at (2.6,-1) {\footnotesize $\times$};
			\node at (2.6,-1.35) {\footnotesize $\rm S^5$};			
			
			
			\foreach \i in {-0.5,0.7,1.65} \draw[red,fill=red] (\i,0) 	circle (2pt);
			\foreach \i in {-0.1,1.4} \draw[blue,fill=blue] (\i,-2) 	circle (2pt);

			\foreach \i in {-1,...,1}{
				\draw[very thick] (1.2*\i+0.1,-0.08) -- (1.2*\i+0.1,0.08) node [anchor=south] {\footnotesize D5};}
			\foreach \i in {-1,1}{
				\draw[thick] (0.9*\i-0.1,-1.92) -- (0.9*\i-0.1,-2.08) node [anchor=north] {\footnotesize NS5};
			}			
		\end{tikzpicture}
	}\hskip 8mm
	\subfigure[][]{
		\begin{tikzpicture}[xscale=0.8,yscale=0.9]
			\shade [right color=3dcolor!100,left color=3dcolor!100] (-0.3,0)  rectangle (0.3,-2);
			
			\shade [ left color=3dcolor! 100, right color=3dcolor! 100] (0.3-0.01,0)  rectangle (2,-2);
			\shade [ right color=3dcolor! 100, left color=3dcolor! 100] (-0.3+0.01,0)  rectangle (-2,-2);
			
			\draw[thick] (-2,0) -- (2,0);
			\draw[thick] (-2,-2) -- (2,-2);
			\draw[thick] (2,-2) -- +(0,2);
			\draw[thick] (-2,-2) -- +(0,2);
			
			\node at (-1,-0.5) {$\Sigma$};
			
			\foreach \i in {-0.5,0.7} \draw[red,fill=red] (\i,0) 	circle (2pt);
			\foreach \i in {-0.1} \draw[blue,fill=blue] (\i,-2) 	circle (2pt);			
			
			\foreach \i in {-1,...,1}{
				\draw[very thick] (1.2*\i+0.1,-0.08) -- (1.2*\i+0.1,0.08) node [anchor=south] {\footnotesize D5};}
			\foreach \i in {-1,1}{
				\draw[thick] (0.9*\i-0.1,-1.92) -- (0.9*\i-0.1,-2.08) node [anchor=north] {\footnotesize NS5};
			}			
		\end{tikzpicture}
	}
	\caption{General form of supergravity duals for interface/defect CFTs (left), boundary CFTs (center), and 3d CFTs (right) with 3d $\mathcal N=4$ (defect) superconformal symmetry. 
		Each (asymptotically local) $\rm AdS_5\times S^5$ region contributes a half space as conformal boundary.
		The D5 and NS5 branes in the brane construction (e.g.\ figs.~\ref{fig:D3NS5-brane}, \ref{fig:D3D5NS5-brane}) are each represented as sources on one of the two boundary components. Their position is determined by the net number of D3-branes ending on the 5-brane. 
		Single-charge BPS geodesics are shown as red and blue dots; their numbers are determined by the numbers of 5-brane groups. The solid green and dashed gray curves show sample families of two-charge BPS geodesics and non-BPS geodesics, respectively. 
		\label{fig:gen-pic}}
\end{figure}
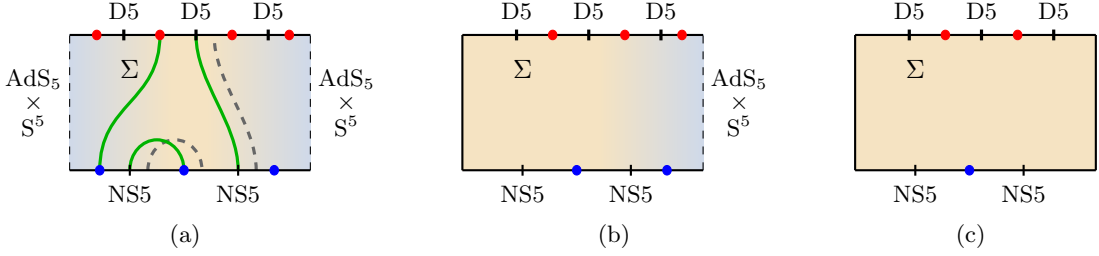

The associated supergravity solutions have a warped product geometry $\rm AdS_4\times S^2\times S^2\times\Sigma$, where $\Sigma$ is a strip.
The general picture is illustrated in fig.~\ref{fig:gen-pic}.
On the upper and lower boundaries of $\Sigma$, an $S^2$ collapses to smoothly close off the 10d geometry. What happens in the horizontal direction depends on the brane setup. For an interface CFT each end is asymptotically locally $\rm AdS_5\times S^5$, with each end contributing one half space as conformal boundary. 
These ends represent semi-infinite D3-branes, e.g.\ as in fig.~\ref{fig:D52NS52-brane}.
For a BCFT the geometry closes off smoothly on one end, leaving one half space as conformal boundary, in line with one stack of semi-infinite D3-branes as in fig.~\ref{fig:D3D5NS5-brane}. For a genuine 3d CFT the geometry closes off on both ends. 
Concrete solutions are specified by a pair of harmonic functions $h_{1/2}$ on $\Sigma$, whose features encode the brane charges.
The 5-branes are represented as distinguished points on the upper and lower boundaries of $\Sigma$, each representing a group of 5-branes with the same net number of D3-branes ending on them;
how many D3-branes end on each 5-brane determines the location of the corresponding source on $\partial\Sigma$.
The $S^2$ which collapses on the boundary of $\Sigma$ with NS5 sources corresponds to $\rm SO(3)_H$; the $S^2$ collapsing on the boundary of $\Sigma$ with D5 sources corresponds to $\rm SO(3)_C$.

To study Penrose limits we determine null geodesics which orbit one or both $S^2$'s, are localized in $\rm AdS_4$ and stationary on $\Sigma$. Such geodesics exist at special points on $\Sigma$. 
We will encounter geodesics whose energy $\mathcal E$ and angular momenta $\mathcal J_i$ on the spheres $S_i^2$ are related as
\begin{align}
	\mathcal E&\geq|\mathcal J_1|+|\mathcal J_2|~.
\end{align}
Geodesics which saturate the inequality will be called BPS. We distinguish single-charge BPS geodesics with $\mathcal J_1=0$ or $\mathcal J_2=0$ and two-charge BPS geodesics with $J_1,J_2\neq 0$.
We will also discuss non-BPS geodesics for which the inequality is strict. 
Points on $\Sigma$ hosting such geodesics can be characterized in terms of the harmonic functions $h_{1/2}$; the derivations are in sec.~\ref{sec:geod-Penrose}, case studies of concrete solutions in sec.~\ref{sec:examples}.
We find the following general picture (illustrated in fig.~\ref{fig:gen-pic}):
\begin{itemize}
	\setlength\itemsep{0.1em}
	\item[--] One single-charge BPS geodesic on each boundary segment of $\Sigma$ connecting a pair of adjacent 5-brane sources, a 5-brane source and an $\rm AdS_5\times S^5$ region, or two $\rm AdS_5\times S^5$ regions.
	\item[--] Families of two-charge BPS geodesics along curves through $\Sigma$ which connect either to the boundary points hosting single-charge BPS geodesics or to 5-brane sources
	\item[--] Families of non-BPS geodesics along certain curves through $\Sigma$ connecting boundary points.
\end{itemize}
Each null geodesic gives rise to a pp-wave sector. The pp-waves are particularly simple for single-charge BPS geodesics and we also give general expressions for the two-charge BPS case.

On each boundary of $\Sigma$ one $S^2$ collapses and geodesics preserve the associated $SO(3)$.
The number of single-charge BPS geodesics at regular boundary points depends only on the number of 5-brane sources, i.e.\ the number of 5-brane groups with the same net number of D3-branes ending on each, not on their location. 
Counting the numbers of single-charge BPS geodesics on the two boundaries of $\Sigma$, which are charged under different $SO(3)$'s, by $C_{1/2}$,
\begin{align}\label{eq:pp-wave-ops-gen}
	C_1&={\rm\#D5} +\begin{cases}
		-1 & \text{3d CFT}\\
		\hphantom{+}0 & \text{BCFT}\\
		+1 & \text{ICFT}
	\end{cases}
	&
	C_2&={\rm\#NS5} +\begin{cases}
		-1 & \text{3d CFT}\\
		\hphantom{+}0 & \text{BCFT}\\
		+1 & \text{ICFT}
	\end{cases}
	.
\end{align}	
where $\#$D5 and $\#$NS5 are the numbers of D5 and NS5 sources, respectively.
Each group of 5-branes with the same net number of D3-branes ending on it contributes a non-Abelian flavor symmetry, and $C_{1/2}$ agree with the  counting of $U(1)$ symmetries in the field theory.
For each single-charge BPS geodesic the pp-wave limit agrees with the $\rm AdS_5\times S^5$ Penrose limit studied in \cite{Berenstein:2002jq}, though the field theory interpretation of the string spectrum is different for different geodesics.

The form of the curves through $\Sigma$ which host families of two-charge BPS geodesics and non-BPS geodesics, on the other hand, depends on the parameters. They may connect points on the same boundary component or on opposing ones, with `phase transitions' between these options as parameters are varied.
The form of the curve determines the range of $\mathcal J_i/\mathcal E$ values covered by the family of geodesics.
The fact that curves hosting two-charge BPS geodesics can end on the 5-brane sources points to additional single-charge BPS geodesics at the 5-brane sources.

\textbf{Outline:} 
The holographic discussion of Penrose limits and BMN-like sectors is split into sec.~\ref{sec:geod-Penrose} for general results and sec.~\ref{sec:examples} for case studies. 
The field theories are discussed in sec.~\ref{sec:fieldtheory}, along with the BMN-like sectors identified holographically. We close with a discussion in sec.~\ref{sec:discussion}.

\section{Penrose limits for \texorpdfstring{$\rm AdS_4\times S^2\times S^2\times\Sigma$}{AdS4xS2xS2xSigma}}\label{sec:geod-Penrose}

In this section we collect general technical results on the Penrose limits.
After reviewing the general $\rm AdS_4\times S^2\times S^2\times\Sigma$ supergravity solutions, we discuss null geodesics and Penrose limits.

The Type IIB supergravity solutions created by intersecting D3/D5/NS5 brane configurations were constructed in \cite{DHoker:2007zhm,DHoker:2007hhe} and \cite{Aharony:2011yc,Assel:2011xz}. The geometry is a warped product ${\rm AdS_4}\times S^2\times S^2\times\Sigma$ with a Riemann surface $\Sigma$. 
The Einstein-frame metric and dilaton take the form
\begin{align}
	ds^2&=f_4^2ds^2_{\rm AdS_4}+f_1^2 ds^2_{S_1^2} +f_2^2ds^2_{S_2^2} +4\rho^2|dz|^2~,
	&
	e^{4\phi}&=\frac{N_2}{N_1}~,
	 \label{eq:met1}
\end{align}
with the dilaton normalized such that $\tau=\chi+ie^{-2\phi}$ \cite{DHoker:2007zhm,DHoker:2007hhe}. The metric functions are
\begin{align}	\label{eq:fs}
	f_4^8&=16\frac{N_1N_2}{W^2}~, & f_1^8&=16h_1^8\frac{N_2 W^2}{N_1^3}~, & f_2^8&=16 h_2^8 \frac{N_1 W^2}{N_2^3}~,
	&
	\rho^8&=\frac{N_1N_2W^2}{h_1^4h_2^4}~,
\end{align}
and we have
\begin{align}
	W&=\partial\bar\partial (h_1 h_2)~, & N_i &=2h_1 h_2 |\partial h_i|^2 -h_i^2 W~.
\end{align}
Concrete solutions are specified by a pair of harmonic functions $h_1, h_2$ on $\Sigma$.
The remaining fields are the dilaton as well as the 3-form and 5-form field strengths, given by
\begin{align}\label{eq:H3F3F5}
	H_{(3)}&=\vol_{S_1^2}\wedge\, db_1~, \qquad\qquad
	F_{(3)}=\vol_{S_2^2}\wedge\, db_2~,
	\nonumber\\
	F_{(5)}&=-4  \vol_{AdS_4}\wedge\, dj_1+4 f_1^2f_2^2f_4^{-4}\vol_{S_1^2}\wedge \vol_{S_2^2}\wedge \star_2 dj_1~,
\end{align}
where $\star_2$ denotes Poincar\'e duality with respect to the metric on $\Sigma$.
$b_1$, $b_2$, $j_1$ are functions on $\Sigma$.
The harmonic functions are split into holomorphic and anti-holormorphic parts as
\begin{align}\label{eq:h12-cA12}
	h_1&=-i (\cA_1-\bar \cA_1)~, & h_2&=\cA_2+\bar \cA_2~,
	\nonumber\\
	h_1^D&=\cA_1+\bar \cA_1~, & h_2^D&=i(\cA_2-\bar \cA_2)~.
\end{align}
Then
\begin{align}\label{eq:b1b2}
	b_1&=\frac{2h_1^2h_2}{N_1}\cY+2h_2^D~,
	&
	b_2&=\frac{2h_1 h_2^2}{N_2}\cY-2h_1^D~,
	&
	j_1&=3\cC+3\bar\cC-3\cD+\frac{h_1h_2}{W}\cY~,
\end{align}
where
$\cY=i(\partial h_1\bar\partial h_2-\bar\partial h_1\partial h_2)$,
$\cD=\bar\cA_1\cA_2+\cA_1\bar\cA_2$ and $\partial\cC=\cA_1\partial\cA_2-\cA_2\partial\cA_1$.
Depending on the choice of $h_{1/2}$ the solutions may describe 4d interface CFTs, boundary CFTs or 3d CFTs. For the solutions we discuss $\Sigma$ can be taken as strip with complex coordinate $z=x+iy$,
\begin{align}\label{eq:Sigma}
	\Sigma&=\left\lbrace z=x+iy\in\CC \ \big\vert \ 0\leq y\leq \frac{\pi}{2}\right\rbrace~.
\end{align}
The functions $h_{1/2}$ are chosen such that one $S^2$ collapses on each boundary to close off the internal space (subject to regularity conditions spelled out in \cite{DHoker:2007hhe}).
Examples will be discussed in sec.~\ref{sec:examples}.

\subsection{Null geodesics}\label{sec:null-geodesics}

We seek null geodesics embedded at the origin of global $\rm AdS_4$ with angular momentum in the internal space, moving along great circles in the $S^2$'s.
At general points of $\Sigma$ adding a geodesic breaks both $SO(3)$ R-symmetries. But at points where one $S^2$ collapses the geodesic can preserve the $SO(3)$ associated with that $S^2$. It can then only have angular momentum on the other $S^2$.

For this section we use the Einstein-frame metric (\ref{eq:met1}), noting that null geodesics are identical in string frame.
We choose explicit coordinates on the $\rm AdS_4$ and $S^2$ spaces as
\begin{align}\label{eq:ads4-global}
	ds^2_{\rm AdS_4}&=
	-\cosh^2\! \hat{\rho}\,dt^2 +d\hat{\rho}^2+\sinh^2\hat{\rho} ds^2_{S^2}~,
	&
	ds^2_{S_i^2}&=d\theta_i^2+\cos^2\!\theta_i\,d\phi_i^2~.
\end{align}
Geodesics at the origin of $\rm AdS_4$ and rotating along great circles in the $S^2_i$ have $\hat\rho=\theta_i=0$.
This leaves $\phi_i(\lambda)$, $t(\lambda)$ and a complex function $z(\lambda)$ with affine parameter $\lambda$. 
We focus on geodesics which are fixed on $\Sigma$.
They are characterized by conserved quantities $\mathcal E$, $\mathcal J_i$ and a point $z_\star$ in $\Sigma$,
\begin{align}\label{eq:BPS-geod}
	\mathcal E&=f_4^2 \frac{dt}{d\lambda}~, & \mathcal J_i&=f_i^2\frac{d\phi_i}{d\lambda}~, & 
	z&=z_\star~.
\end{align}
The general solution for this ansatz is discussed in app.~\ref{app:geod-details}.
In anticipation of the field theory BPS conditions in (\ref{eq:3dN4-multiplets}), we distinguish BPS geodesics with
\begin{align}\label{eq:BPS-ab}
	|\mathcal J_1|+|\mathcal J_2|&=\mathcal E~,
\end{align}
and non-BPS geodesics where this condition is not satsified.
The task then is to determine the allowed $z_\star$. The results of  app.~\ref{app:geod-details} lead to the following cases:
\begin{enumerate}
	\item BPS geodesics on the boundary of $\Sigma$ where $S^2_1$ collapses spin only on $S_2^2$.
	They have $\mathcal J_1=0$ and $\mathcal E=|\mathcal J_2|$. They need $z_\star$ s.t.
	\begin{align}\label{eq:nice-cond-1}
		h_1&=0~, &  \partial h_2&=0~.
	\end{align}
	For regular $\rm AdS_4\times S^2\times S^2\times \Sigma$ solutions
	$h_2$ satisfies a Neumann boundary condition on the boundary where $h_1=0$ \cite[sec.~3.5]{DHoker:2007hhe}.
	So $\partial h_2=0$ reduces to the requirement that the derivative of $h_2$ along the boundary vanishes, i.e.\ $z_\star$ is an extremum of $h_2$ seen as function of a real coordinate along the boundary of $\Sigma$. We expect isolated points as solutions.
	
	\item  BPS geodesics on the boundary of $\Sigma$ where $S^2_2$ collapses spin only on $S_1^2$.
	They have $\mathcal J_2=0$ and $\mathcal E=|\mathcal J_1|$, and need $z_\star$ s.t.
	\begin{align} \label{eq:nice-cond-2}
		h_2&=0~, &  \partial h_1&=0~.
	\end{align}
	Analogously to case 1, this singles out extrema of $h_1$ along the boundary of $\Sigma$ where $h_2=0$.
	
	\item BPS geodesics in the interior of $\Sigma$, spinning on $S_1^2$ and $S_2^2$ such that $\mathcal E=|\mathcal J_1|+|\mathcal J_2|$, need $z_\star$ such that
	\begin{align} \label{eq:h1h2-cond-gen}
		\partial h_1\bar\partial h_2-\bar\partial h_1\partial h_2&=0~.
	\end{align}
	This is one real condition for the complex $z_\star$, so we may expect solutions along curves in $\Sigma$. In this case  $\mathcal J_i^2/\mathcal E^2=f_i^4/f_4^4$,
	with convenient expressions for $f_i^2/f_4^2$ given in (\ref{eq:f12df4}).
	
	\item Further geodesics, which are not necessarily BPS, can be placed at points $z_\star$ solving
	\begin{align}\label{eq:geod-gen}
		\partial\left(\frac{h_2\partial h_1}{h_1\partial h_2}\right)\bar\partial\left(\frac{h_2\bar\partial h_1}{h_1\bar\partial h_2}\right)-\bar\partial\left(\frac{h_2\partial h_1}{h_1\partial h_2}\right)\partial\left(\frac{h_2\bar\partial h_1}{h_1\bar\partial h_2}\right)&=0~,
	\end{align}
	subject to the constraint that $\mathcal J_i^2/\mathcal E^2\geq 0$, where
	\begin{align}\label{eq:J12-non-BPS}
		\frac{\mathcal J_1^2}{\mathcal E^2}&=\frac{f_1^4}{f_4^4}\frac{\partial_z(f_4^2/f_2^2)}{\partial_z(f_1^2/f_2^2)}~,
		&
		\frac{\mathcal J_2^2}{\mathcal E^2}&=\frac{f_2^4}{f_4^4}\frac{\partial_z(f_4^2/f_1^2)}{\partial_z(f_2^2/f_1^2)}
		~.
	\end{align}
\end{enumerate}
How many points on $\Sigma$ hosting such geodesics exist depends on $h_{1/2}$.
Cases 1, 2 can be seen as limits of case 3 when $z_\star$ approaches a boundary.
The geodesics then preserve the $SO(3)$ associated with the collapsed $S^2$. Case 4 also becomes single charge if $z_\star$ approaches a boundary of $\Sigma$.

We now show a number of statements which determine the number of solutions to the conditions stated above and the nature of singular points in the solutions.

\textbf{Extrema of $h_{1/2}$ along $\partial\Sigma$ are minima:} 
We focus on the boundary where $h_2=0$, the case $h_1=0$ follows analogously.
We can locally take the boundary where $h_2=0$ for generic $\Sigma$ as a segment of the real line, $y=0$, with the interior of $\Sigma$ at $y>0$. At this boundary we have 
\begin{align}
	\frac{f_4^2}{\rho^2}&=\left\vert\frac{2h_1h_2}{W}\right\vert
	~,
	&
	f_4^6\rho^2&=16h_1^2|\partial h_2|~.
\end{align}
$f_4^2$ and $\rho^2$ are finite and non-zero on this boundary by assumption, and so is $h_1$.
So $W/h_2$ is finite and non-zero from the first expression, and from the second expression $\partial h_2\neq 0$.
We then use
\begin{align}
	\partial_y W\vert_{y=0}&=
	\frac{1}{4} \left(\partial_x h_1\partial_x\partial_y h_2-\partial_x^2h_1\partial_yh_2 \right)\big\vert_{y=0}~.
\end{align}
At a point $x_\star$ where $\partial_x h_1\vert^{}_{x=x_\star}=0$, the first term vanishes. Since $h_2$ has to be positive in the interior of $\Sigma$ \cite[sec.~3.5]{DHoker:2007hhe}, we have $\partial_y h_2\geq 0$.
Since $\partial h_2\neq 0$ the inequality is strict, $\partial_y h_2> 0$.
Since $W/h_2$ is finite, we also have $\partial_y W\neq 0$.
Since $W$ is negative in the interior, we have $\partial_y W< 0$. We thus need $\partial_x^2h_1>0$. That is, all extrema of $h_1(x,0)$ are minima.

\textbf{Counting BPS geodesics on $\partial\Sigma$:}
What does this imply for points hosting BPS geodesics on the boundary of $\Sigma$? At 5-brane poles and in asymptotic $\rm AdS_5\times S^5$ regions $h_1$ diverges (e.g.\ \cite[eq.~(3.7)]{Aharony:2011yc} and \cite[eq.~(2.23)]{Aharony:2011yc}). In-between $h_1(x,0)$ is a smooth function. With only minima as extrema, there is precisely one extremum between any two points where $h_1\rightarrow\infty$. As a result:

There is exactly one solution to (\ref{eq:nice-cond-1}), (\ref{eq:nice-cond-2}) on the boundary of $\Sigma$ between adjacent 5-brane poles, between a 5-brane pole and an $\rm AdS_5\times S^5$ region, and between adjacent $\rm AdS_5\times S^5$ regions.

\medskip

\textbf{Singular points:} In addition to the regular points on $\Sigma$, there are singular points in the form of 5-brane sources and asymptotic $\rm AdS_5\times S^5$ regions. One can ask if the conditions for BPS geodesics are satisfied in a limiting sense at these points. We find:
\begin{enumerate}
	\item[--] 5-brane sources satisfy the conditions (\ref{eq:bps-null-geod}) in a limiting sense:
	Depending on the type of 5-brane source,
	$f_4^2$ has the same behavior as either $f_1^2$ or $f_2^2$,
	as can be seen in \cite[eq.~(3.10)]{Aharony:2011yc}.
	
	\item[--] Where an $\rm AdS_5\times S^5$ region emerges, $f_4^2\,{\rightarrow}\,\infty$ with $f_{1/2}^2$ finite, (\ref{eq:bps-null-geod}) is not satisfied. Likewise when the geometry closes off with both $S^2$'s collapsing, $f_{1/2}^2\rightarrow 0$ with $f_4^2$ finite.
\end{enumerate}
One may resolve a source corresponding to $N_5$ 5-branes into $N_5$ separated sources for single 5-branes. This would lead to $N_5-1$ BPS geodesics at regular points. The resolution changes the brane configuration and field theory, but viewing a source of charge $N_5$ as limiting case of $N_5$ unit-charge sources suggests that the former may host additional single-charge BPS geodesics.

\textbf{Connection to Wilson loops:} The condition for BPS geodesics in (\ref{eq:nice-cond-1}) matches the BPS condition for fundamental strings representing Wilson loops  \cite[eq.~(4.12)]{Coccia:2021lpp}, and (\ref{eq:nice-cond-2}) matches the corresponding condition for vortex loops.
The counting of BMN sectors thus matches the counting of `regular' fundamental Wilson and vortex loops \cite{Coccia:2021lpp}. 
For the latter there are additional string embeddings at the poles. They can be understood as limiting cases of Wilson loops in large antisymmetric representations, represented by probe branes in the interior of $\Sigma$, and their features can be understood precisely from the matrix models (discussion around \cite[eq.~(3.11)]{Coccia:2021lpp}, \cite{Uhlemann:2020bek}).

\subsection{Penrose limits}\label{sec:Penrose}
For the Penrose limit (in string frame) we start with cases 1, 2.
We give expressions for case 1; case 2 follows by exchanging the $S^2$'s and the harmonic functions and inverting $e^{\phi}$.
Derivations are in app.~\ref{app:geod-details}.
Without loss of generality we take the boundary locally along $y=0$ on $\Sigma$. 
For the Penrose limit we replace the coordinates as follows,
\begin{align}\label{eq:Penrose-dSigma-coord}
	t&=x^+ + \frac{x^-}{\tilde f_4^2}~,&
	\hat{\rho}&=\frac{r}{\tilde f_4}~,&
	\theta_2&=\frac{x_6}{\tilde f_4}~,&
	\phi_2&=x^+ - \frac{x^-}{\tilde f_4^2}~,&
	z&=z_\star+\frac{x_7+ix_8}{2 \tilde\rho}~.
\end{align}
In this expression the string-frame metric functions $\tilde f_i^2=e^\phi f_i^2$, $\tilde\rho^2=e^{\phi}\rho^2$ are evaluated at $z_\star$.
They are large in the holographic limit. The coordinates of the $S^2$ in $\rm AdS_4$ and the $S_1^2$ in the internal space are not rescaled.
The Penrose limit of the string-frame metric becomes
\begin{align}\label{eq:ppwave-metric-dSigma}
	ds^2&=- 4 dx^+ dx^- + dr^2 + r^2 ds^2_{S^2}+ dx_6^2+ dx_7^2 + dx_8^2 + x_8^2 ds^2_{S_1^2}-
	(dx^+)^2 \left(r^2 + x_6^2 + x_7^2 + x_8^2\right).
\end{align}
$(r,ds^2_{S^2})$ and $(x^8, ds^2_{S_1^2})$ each parametrize an $\RR^3$ in spherical coordinates. 
Moreover,
\begin{align}\label{eq:Penrose-dSigma-dilaton}
	e^{+4\phi}&=-\frac{h_2\partial_y^2 h_2}{(\partial_y h_1)^2}~,
	&
	H_{(3)}&=F_{(3)}=0~, & 
	F_{(5)}&=(\mathds{1}+\star)4 r^2  dr\wedge dx^+ \wedge \vol_{S^2}\wedge dx_7  ~.
\end{align}
The functions $h_{1/2}$ only enter in specifying the (constant) dilaton.
This pp-wave limit has the same form as the one identified by BMN for $\rm AdS_5\times S^5$ in \cite{Berenstein:2002jq}.
$\rm AdS_5\times S^5$ emerges for a particular choice of $h_{1/2}$, discussed in sec.~\ref{sec:AdS5xS5}, but this result for the pp-wave limit applies for arbitrary $h_{1/2}$.
The number of points on $\Sigma$ which solve (\ref{eq:nice-cond-1}), (\ref{eq:nice-cond-2}) and lead to this pp-wave limit depends on $h_{1/2}$.

To characterize the operators described by the Penrose limit we follow \cite{Berenstein:2002jq}. With energy $\Delta=i\partial_t$ and angular momentum $J_2=-i\partial_{\phi_2}$ we have
\begin{align}\label{eq:Penrose-nice-momenta}
	-p_+&=i\partial_{x^+}=i(\partial_t+\partial_{\phi_2})=\Delta-J_2~,
	\nonumber\\
	-p_-&=i\partial_{x^-}=\frac{1}{\tilde f_4^2}i(\partial_t-\partial_{\phi_2})=\frac{\Delta+J_2}{\tilde f_4^2}~.
\end{align}
The BPS bound $\Delta\geq |J_2|$ keeps $-p_{\pm}$ non-negative. The light cone Hamiltonian is $H_{\rm LC}=-p_+$ and states with $H_{\rm LC}=0$ have $\Delta=J_2$. States with $p_\pm=-2p^{\mp}$ of order one correspond to $\Delta,J_2\sim \tilde f_4^2$.

The string spectrum on the pp-wave solution can be taken from \cite[(3.2),(3.6)]{Berenstein:2002jq}.
There are 8 bosonic and 8 fermionic worldsheet fields, whose quantization leads to
\begin{align}
	H_{\rm LC}=-p_+&=\sum_{n=-\infty}^{+\infty}N_n\sqrt{1+\frac{n^2}{(\alpha'p^+)^2}}~,&
	P=\sum_{n=-\infty}^{+\infty}nN_n&\stackrel{!}{=}0~,
\end{align}
where $n$ is the momentum, with $n>0$ ($n<0$) for left (right) movers and $n=0$ included, and $N_n$ is the occupation number including bosons and fermions.
Following the arguments in \cite{Berenstein:2002jq} for the translation to $\Delta$, $J$ using (\ref{eq:Penrose-nice-momenta}) and generalizing the expressions to cover cases 1,2 together leads to
\begin{align}\label{eq:string-spectrum}
	\Delta-J_i&=\sum_{n=-\infty}^\infty N_n (\Delta-J_i)_n~, &
	(\Delta-J_i)_n&=\sqrt{1+\frac{n^2\tilde f_4^4}{{\alpha'}^2J_i^2}}~.
\end{align}
The choice of supergravity solution and dual field theory only enters through the string frame warp factor $\tilde f_4^4=e^{2\phi}f_4^4$. 
At points where (\ref{eq:nice-cond-1}) is satisfied $\tilde f_4^2=2h_2$.

We briefly discuss case 3, i.e.\ geodesics in the interior of $\Sigma$ at points satisfying (\ref{eq:h1h2-cond-gen}). 
The pp-wave geometry can be obtained from the ansatz in (\ref{eq:Penrose-gen-transf}), which in particular leads to 
\begin{align}
t&=x^+ +\frac{x^-}{\tilde f_4^2}~, &
\phi_1&=x^+ -\frac{x^-}{\tilde f_4^2} +\frac{\tilde f_2}{\tilde f_1\tilde f_4} x^9~,&
\phi_2&=x^+ -\frac{x^-}{\tilde f_4^2}- \frac{\tilde f_1}{\tilde f_2\tilde f_4} x^9~.
\end{align}
For this case the pp-wave geometry depends on the functions $h_{1/2}$. 
The explicit form is given in \eqref{eq:Penrose-gen-ds2}.
The momenta become, with $J_i=-i\partial_{\phi_i}$,
\begin{align}\label{eq:Penrose-momenta}
	-p_+&=i\partial_{x^+}=i(\partial_t+\partial_{\phi_1}+\partial_{\phi_2})=\Delta-J~,
	&
	J\equiv J_1+J_2~,
	\nonumber\\
	-p_-&=i\partial_{x^-}=\frac{1}{\tilde f_4^2}i(\partial_t-\partial_{\phi_1}-\partial_{\phi_2})=\frac{\Delta+J}{\tilde f_4^2}~,
	\nonumber\\
	+p_9&=i\partial_{x^9}=\frac{i}{\tilde f_4}\frac{\tilde f_2^2\partial_{\phi_1}-\tilde f_1^2\partial_{\phi_2}}{\tilde f_1\tilde f_2}=\frac{\tilde f_1^2J_1-\tilde f_2^2J_2}{\tilde f_4\tilde f_1\tilde f_2}~.
\end{align}
The BPS bound $\Delta\geq J$ makes $-p_\pm$ non-negative and states with vanishing light cone Hamiltonian have $\Delta=J$.
Finite $p_-$ amounts to $\Delta,J_1,J_2\sim \tilde f_4^2$.
States with vanishing $p_9$ have (using (\ref{eq:bps-null-geod}))
\begin{align}\label{eq:Penrose-J1J2}
	\frac{J_1}{J}&=\frac{f_1^2}{f_4^2}\Big\vert_{z=z_\star}~,
	&
	\frac{J_2}{J}&=\frac{f_2^2}{f_4^2}\Big\vert_{z=z_\star}~.
\end{align}
The pp-wave geometry in \eqref{eq:Penrose-gen-ds2} is not as simple as for the single-charge case.
We leave a more detailed discussion of the two-charge and non-BPS cases for the future.

\section{Holographic case studies}\label{sec:examples}

In this section we discuss the holographic duals for a sample of 4d $\mathcal N=4$ SYM boundary and interface CFTs, realized by particular choices for the functions $h_{1/2}$ in the solutions of sec.~\ref{sec:geod-Penrose}, and use the results of sec.~\ref{sec:geod-Penrose} to study Penrose limits.
We begin with $\rm AdS_5\times S^5$ and the Janus interface, and then move on to theories
engineered by D3-branes ending on or intersecting 5-branes.

\subsection{\texorpdfstring{$\rm AdS_5\times S^5$ and Janus interface}{AdS5xS5 and Janus interface}}\label{sec:AdS5xS5}

\begin{figure}
	\begin{tikzpicture}
		\draw[dashed] (0,0) -- (0,2);
		\fill [color=blue!30] (-2.5,0) rectangle (0,2);
		\fill [color=red!30] (0,0) rectangle (2.5,2);
		\node at (-1.25,1.3) {$\mathcal N=4$ SYM};
		\node at (-1.25,0.7) {$g^{}_{\rm YM,L}$};
		
		\node at (1.25,1.3) {$\mathcal N=4$ SYM};
		\node at (1.25,0.7) {$g^{}_{\rm YM,R}$};
	\end{tikzpicture}
	\hskip 20mm
	\begin{tikzpicture}[scale=1]
		\shade [right color=3dcolor!100,left color=3dcolor!100] (-0.3,0)  rectangle (0.3,-2);
		
		\shade [ left color=3dcolor! 100, right color=red!30] (0.3-0.01,0)  rectangle (2,-2);
		\shade [ right color=3dcolor! 100, left color=blue!30] (-0.3+0.01,0)  rectangle (-2,-2);

		\draw[thick] (-2,0) -- (2,0);
		\draw[thick] (-2,-2) -- (2,-2);
		\draw[dashed] (2,-2) -- +(0,2);
		\draw[dashed] (-2,-2) -- +(0,2);
		
		\node at (-0.5,-0.5) {$\Sigma$};
		\node at (2.6,-0.65) {\footnotesize $\rm AdS_5$};
		\node at (2.6,-1) {\footnotesize $\times$};
		\node at (2.6,-1.35) {\footnotesize $S^5$};			
		
		\node at (-2.6,-0.65) {\footnotesize $\rm AdS_5$};
		\node at (-2.6,-1) {\footnotesize $\times$};
		\node at (-2.6,-1.35) {\footnotesize $S^5$};						
	\end{tikzpicture}
	\caption{Left: Janus interface between $\mathcal N=4$ SYM on two half spaces with independent couplings. Right: Schematic form of the $\rm AdS_4\times S^2\times S^2\times\Sigma$ supergravity dual. 
	Two locally $\rm AdS_5\times S^5$ regions emerge at $\Re(z)\rightarrow\pm\infty$, each contributing an $\rm AdS_4$ conformal boundary which is conformally equivalent to a half space. The dilaton takes independent values at the two ends of $\Sigma$.
	$\rm AdS_5\times S^5$ arises as special case $g^{}_{\rm YM,L}=g^{}_{\rm YM,R}$.
	\label{fig:Janus-sugra}}
\end{figure}
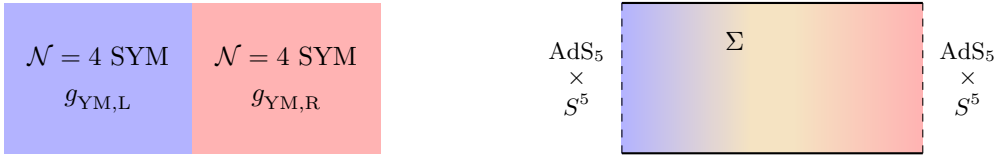

We start with the $\rm AdS_5\times S^5$ solution of Type IIB, dual to standard 4d $\mathcal N=4$ SYM, which arises as a special case of the general $\rm AdS_4\times S^2\times S^2\times\Sigma$ solutions. The harmonic functions are
\begin{align}\label{eq:h12-AdS5S5}
	h_1&=-i \frac{R^2}{4\sqrt{g_s}}(\sinh z -\sinh\bar z)~, & h_2&=\frac{\sqrt{g_s}R^2}{4}(\cosh z+\cosh \bar z)~.
\end{align}
With real coordinates $z=x+iy$, the Einstein-frame metric (\ref{eq:met1}) and dilaton become
\begin{align}\label{eq:AdS5S5-metric}
	ds^2&=R^2\left[\cosh^2\!x\,ds^2_{\rm AdS_4}+\sin^2\!y\,ds^2_{S^2_1}+\cos^2\!y\,ds^2_{S^2_2}+dx^2+dy^2\right],
	&
	e^{4\phi}&=g_s^2~.
\end{align}
The $x$-direction combines with ${\rm AdS_4}$ to form $\rm AdS_5$, while the $y$-direction combines with the $\rm S^2$'s to form $\rm S^5$.
The Einstein-frame radius is $R^4=4\pi{\alpha^\prime}^2N$, with $N$ the number of D3-branes.
For this choice of $h_{1/2}$ the bosonic symmetry of the solution enhances from $SO(2,3)\times SO(3)\times SO(3)$ to $SO(2,4)\times SO(6)$, and the 16 supersymmetries of the general solutions enhance to 32.
The coordinates only make the part of the symmetry group shared by all solutions manifest.

We find two single-charge BPS geodesics from (\ref{eq:nice-cond-1}), (\ref{eq:nice-cond-2}): at $x=0$ with $y\in(0,\frac{\pi}{2})$.
These are equivalent to the geodesics studied in \cite{Berenstein:2002jq}.\footnote{Upon embedding S$^5$ in $\RR^6$ with a parametrization leading to the coordinates in (\ref{eq:AdS5S5-metric}), the angular momenta on the two $S^2$'s become angular momenta in orthogonal $\RR^2$ planes.}
For the associated pp-wave solutions the string spectrum is given by (\ref{eq:string-spectrum}) with $\tilde f_4^4(z_\star)=g_sR^4$, where $4\pi g_s=g_{\rm YM}^2$. This recovers the results of \cite{Berenstein:2002jq}.

The condition (\ref{eq:h1h2-cond-gen}) for two-charge BPS geodesics fixed on $\Sigma$ and spinning on both $S^2$'s in accordance with the BPS condition (\ref{eq:BPS-ab}) leads to
\begin{align}\label{eq:AdS5xS5-geod}
	z_\star&=iy~, \qquad 0\leq y\leq \frac{\pi}{2}~.
\end{align}
A BPS null geodesic can be placed at each of these points.
The angular momenta for a geodesic at fixed $y$
correspond to the charges of the associated operator via (\ref{eq:Penrose-J1J2}), and are given by
\begin{align}\label{eq:J1J2-AdS5S5}
	\frac{J_1}{J}&=\sin^2y~,
	&
	\frac{J_2}{J}&=\cos^2y~.
\end{align}
We recover $J_1=0$ at $y=0$ and $J_2=0$ at $y=\frac{\pi}{2}$. 
Penrose limits for such two-charge geodesics in $\rm AdS_5\times S^5$ were studied in \cite{Grignani:2009ny, Bertolini:2002nr} using an $S^1\times S^3$ slicing of $S^5$ with two angular momenta on $S^3$.

Relaxing the BPS condition does not lead to additional non-BPS null geodesics from (\ref{eq:geod-gen}). But for this solution we expect additional null geodesics which move on $\Sigma$, which are not captured.

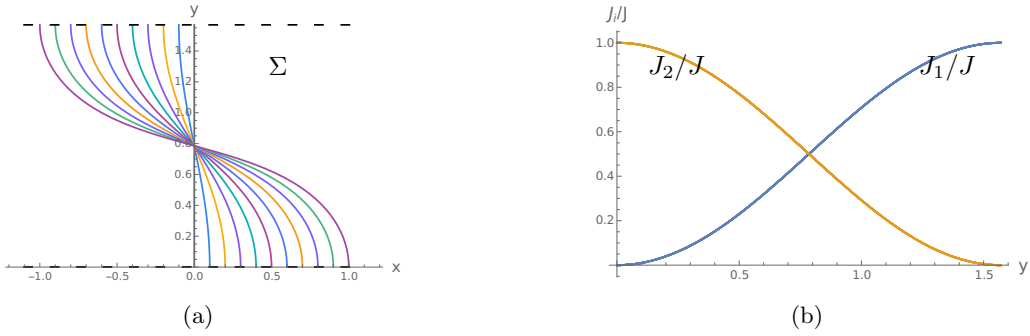
\begin{figure}
	\subfigure[][]{\label{fig:Janus-curves}
		\begin{tikzpicture}
		\node at (0,0) {\includegraphics[width=0.32\linewidth]{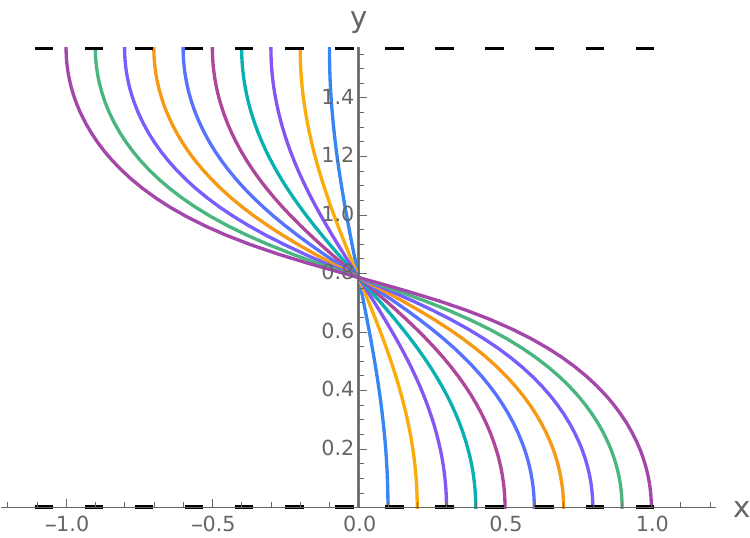}};
		\node at (1,1) {$\Sigma$};
		\end{tikzpicture}
	}
	\hskip 20mm
	\subfigure[][]{\label{fig:Janus-J12}
		\begin{tikzpicture}
			\node at (0,0) {\includegraphics[width=0.35\linewidth]{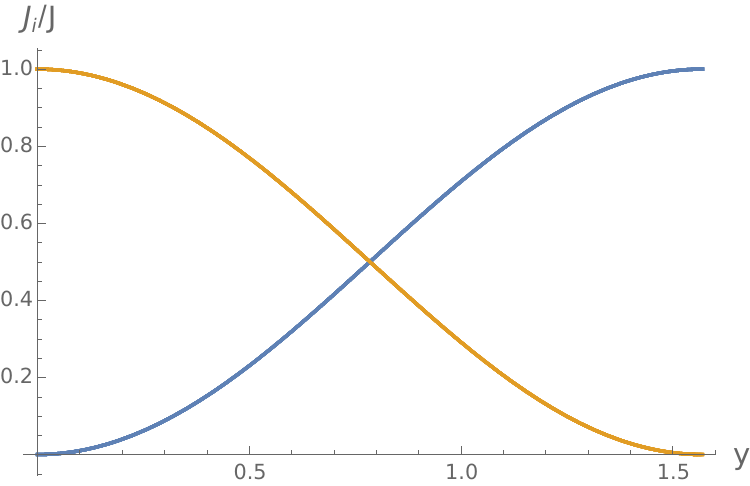}};
			\node at (1.8,1){$J_1/J$};
			\node at (-1.8,1){$J_2/J$};
		\end{tikzpicture}
	}
	\caption{Left: curves along which BPS null geodesics can be placed for $\delta=\lbrace \frac{1}{10},\frac{2}{10},\ldots,1\rbrace$. Curves for $\delta<0$ can be obtained by a vertical reflection. Right: $J_i/J$ for geodesics at $\Im(z)=y$ (independent of $\delta$).}
\end{figure}

\bigskip

\textbf{Janus interface:} The supersymmetric Janus interface separates two 4d $\mathcal N=4$ SYM theories on half spaces with identical gauge groups but different couplings.
The preserved R-symmetry is $SO(3)\times SO(3)$, as discussed in sec.~\ref{sec:Janus-QFT}.
The $\rm AdS_4\times S^2\times S^2\times \Sigma$ dual was  discussed in \cite[sec.~10.3]{DHoker:2007zhm}.
In this solution the symmetry is not enhanced to $SO(2,4)\times SO(6)$.
The harmonic functions are
\begin{align}\label{eq:h12-Janus}
	h_1&=-i\frac{R^2}{4\sqrt{g_0\cosh(2\delta)}}\sinh(z+\delta)+\rm{c.c.}
	&
	h_2&=\frac{R^2\sqrt{g_0}}{4\sqrt{\cosh(2\delta)}}\cosh(z-\delta)+\rm{c.c.}
\end{align}
with a real parameter $\delta$. For $\Re(z)\rightarrow\pm\infty$ the solution asymptotes to $\rm AdS_5\times S^5$ with curvature radius $R$. The dilaton is position-dependent on $\Sigma$ with asymptotic values
\begin{align}\label{eq:Janus-couplings}
	\lim_{\Re(z)\rightarrow\pm\infty}e^{4\phi}=g_0^2e^{\mp4\delta}~.
\end{align}
These asymptotic dilaton values control the 4d $\mathcal N=4$ SYM gauge couplings on the two half spaces,
\begin{align}
	g_{\rm YM,L}^2&=4\pi g_0e^{-2\delta}~, & g_{\rm YM,R}^2&=4\pi g_0e^{+2\delta}~.
\end{align}
Their difference is controlled by $\delta$.
For $\delta=0$ the solution reduces to ${\rm AdS}_5\times S^5$.

The conditions (\ref{eq:nice-cond-1}), (\ref{eq:nice-cond-2}) again yield 2 single-charge BPS geodesics on the boundary of $\Sigma$, with
\begin{align}\label{eq:Janus-geod-boundary}
	z&=\delta:\quad  \tilde f_4^4=R^4g_0\sech(2\delta)~,
	&
	z&=-\delta+\frac{i\pi}{2}:\quad  \tilde f_4^4=R^4g_0\cosh(2\delta)~.
\end{align}
As shown in sec.~\ref{sec:Penrose}, the Penrose limit for each leads to a maximally symmetric pp-wave solution, just as in $\rm AdS_5\times S^5$, despite the Janus solution having less symmetry.
However, the Janus deformation changes $\tilde f_4^2$ at the locations of the two geodesics, and thus the spectra of operators described by the pp-wave limit via (\ref{eq:string-spectrum}), which leads to (\ref{eq:Janus-spectrum}) below. The individual pp-wave limits are maximally symmetric but the Janus deformation lifts the degeneracy between them.

The condition for two-charge BPS geodesics (\ref{eq:h1h2-cond-gen}) again leads to a single curve in $\Sigma$ which interpolates between the two single-charge geodesics on opposite boundary components,
\begin{align}\label{eq:Janus-curve}
	\sinh (2 x)&=\sinh (2 \delta ) \cos (2 y)~.
\end{align}
This is shown in fig.~\ref{fig:Janus-curves}. 
The charges $J_{i}/J$ determined from (\ref{eq:Penrose-J1J2}) agree with the $\rm AdS_5\times S^5$ result (\ref{eq:J1J2-AdS5S5}), shown in fig.~\ref{fig:Janus-J12},
and likewise interpolate between the charges of the single-charge geodesics.
As for $\rm AdS_5\times S^5$, no additional non-BPS null geodesics arise from  (\ref{eq:geod-gen}).

\subsection{D3/NS5 and D3/D5 BCFTs}\label{eq:D3NS5-sugra}

\begin{figure}
	\subfigure[][]{\label{fig:D3NS5-sol}
		\begin{tikzpicture}[scale=1]
			\shade [right color=3dcolor!100,left color=3dcolor!100] (-0.3,0)  rectangle (0.3,-2);
			
			\shade [ left color=3dcolor! 100, right color=4dcolor! 100] (0.3-0.01,0)  rectangle (2,-2);
			\shade [ right color=3dcolor! 100, left color=4dcolor! 100] (-0.3+0.01,0)  rectangle (-2,-2);

			\draw[thick] (-2,0) -- (2,0);
			\draw[thick] (-2,-2) -- (2,-2);
			\draw[dashed] (2,-2) -- +(0,2);
			\draw[thick] (-2,-2) -- +(0,2);
			
			\node at (-0.5,-0.5) {$\Sigma$};
			\node at (2.6,-0.65) {\footnotesize $\rm AdS_5$};
			\node at (2.6,-1) {\footnotesize $\times$};
			\node at (2.6,-1.35) {\footnotesize $S^5$};			
			
			\draw[very thick] (0,-2+0.08) -- (0,-2-0.08) node [anchor=north] {\footnotesize NS5};
		\end{tikzpicture}
	}\hskip 3mm
	\subfigure[][]{\label{fig:D3NS5-brane}
		\begin{tikzpicture}[y={(0cm,1cm)}, x={(0.707cm,0.707cm)}, z={(1cm,0cm)}, scale=1.1]
			\draw[gray,fill=gray!100] (0,0,0) ellipse (1.5pt and 3pt);
			
			\foreach \i in {-0.05,0,0.05}{ \draw[thick] (0,-1,\i) -- (0,1,\i);}
			
			\foreach \i in  {-0.075,-0.045,-0.015,0.015,0.045,0.075}{ \draw (0,1.4*\i,0) -- (0,1.4*\i,2.8);}			
			
			\node at (-0.2,-0.2,2.2) {\scriptsize $N_5K$ D3};
			\node at (0,-1.25) {\footnotesize $N_5$ NS5};
		\end{tikzpicture}
	}\hskip 7mm
	\subfigure[][]{\label{fig:D3NS5-brane-2}
		\begin{tikzpicture}
			\foreach \i in {0,1,2,3}{\draw[thick] (\i,1) -- +(0,-2) node[yshift=-3mm] {\footnotesize NS5};}
			\foreach \i in {-1,0,1} {\draw (0,{0.05*\i}) -- +(1,0);}
			\foreach \i in {-2.5,-1.5,-0.5,0.5,1.5,2.5} {\draw (1,{0.05*\i}) -- +(1,0);}
			\foreach \i in {-4,...,4} {\draw (2,{0.05*\i}) -- +(1,0);}
			\foreach \i in {-5.5,-4.5,...,5.5} {\draw (3,{0.05*\i}) -- +(1.5,0);}			
		\end{tikzpicture}		
	}
	\caption{Schematic form of the supergravity dual for the D3/NS5 BCFT (left) and the brane construction (center). 
	Separating the NS5-branes makes the UV gauge theory  (\ref{eq:D3NS5-quiver}) manifest (right, for $N_5=4$, $K=3$).\label{fig:D3NS5-brane-sol}}
\end{figure}
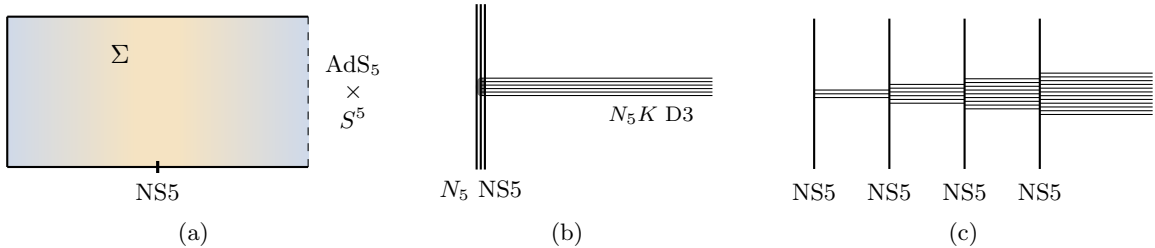

As first example of a BCFT we consider $N_5K$ D3-branes ending on $N_5$ NS5-branes (fig.~\ref{fig:D3NS5-brane}). Each NS5-brane has a net number of $K$ D3-branes ending on it. 
The field theories will be discussed in sec.~\ref{sec:D3NS5-ops}.
The harmonic functions for the supergravity duals are
\begin{align}\label{eq:h1h2-D3NS5-BCFT}
	h_1&=-\frac{i\pi \alpha^\prime}{4} K e^z+\rm{c.c.}
	&
	h_2&=\frac{\pi\alpha^\prime}{4}K e^{z+2\phi_0}-\frac{\alpha^\prime}{4}N_5\ln\tanh\left(\frac{z}{2}\right)+\rm{c.c.}
\end{align}
The pole in $\partial h_2$ at $z=0$ represents the NS5-branes.  A (locally) $\rm AdS_5\times S^5$ region only emerges at $\Re(z)\rightarrow+\infty$, at $\Re(z)\rightarrow-\infty$ the geometry closes off.
The asymptotic dilaton at $\Re(z)\rightarrow\infty$ is
\begin{align}\label{eq:D3NS5-sol-dilaton}
	\lim_{\Re(z)\rightarrow\infty}e^{2\phi}&=e^{2\phi_0}~.
\end{align}
The solution is illustrated in fig.~\ref{fig:D3NS5-sol}.
In the convention of \cite{DHoker:2007zhm,DHoker:2007hhe}, $\tau=\chi+i e^{-2\phi}$; identification with $\tau =\theta/(2\pi)+4\pi i/g_{4d}^2$ in 4d $\mathcal N=4$ SYM leads to $g_{4d}^2=4\pi e^{2\phi_0}$.
The 4d `t Hooft coupling $\lambda$ then is
\begin{align}
	\lambda_{4d}&= N_5K g_{4d}^2=4\pi N_5K e^{2\phi_0}~.
\end{align}

\begin{figure}
	\centering
	\subfigure[][]{\label{fig:D3NS5curves}
		\begin{tikzpicture}
		\node at (0,0) {\includegraphics[width=0.31\linewidth]{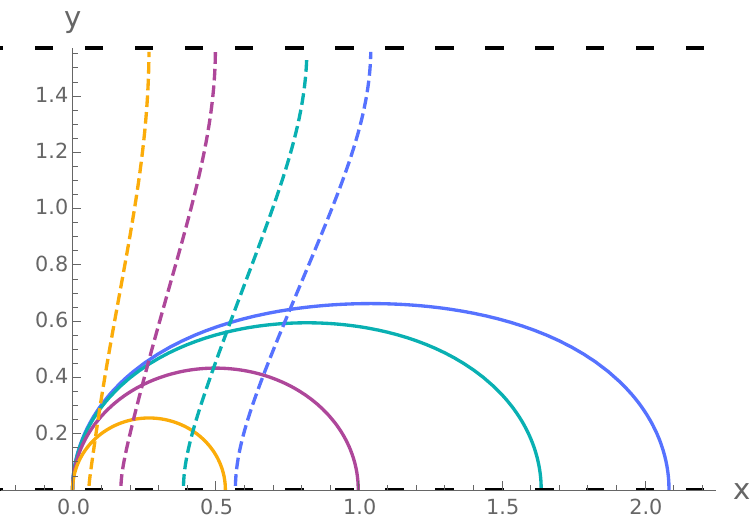}};
		\node at (1,1) {$\Sigma$};
		\end{tikzpicture}
	}
	\hskip -4mm
	\subfigure[][]{\label{fig:D3NS5-J12}
		\begin{tikzpicture}
			\node at (0,0) {
		\includegraphics[width=0.32\linewidth]{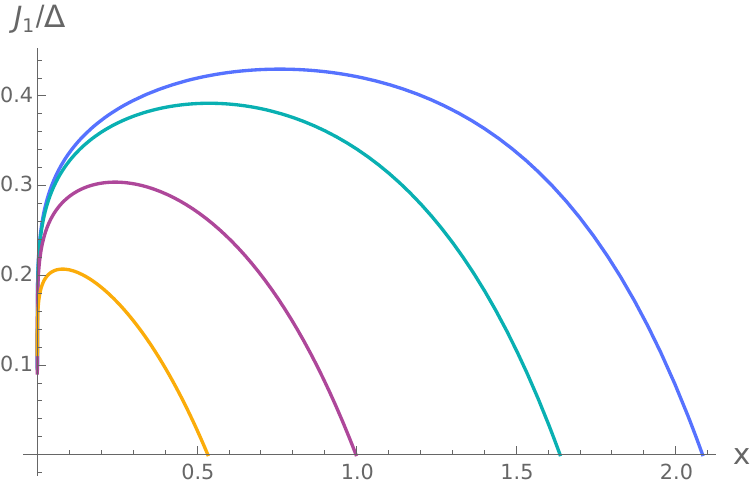}
		};
		\end{tikzpicture}
	}
	\hskip -4mm
	\subfigure[][]{\label{fig:D3NS5-J1max}
		\begin{tikzpicture}
			\node at (0,0) {\includegraphics[width=0.31\linewidth]{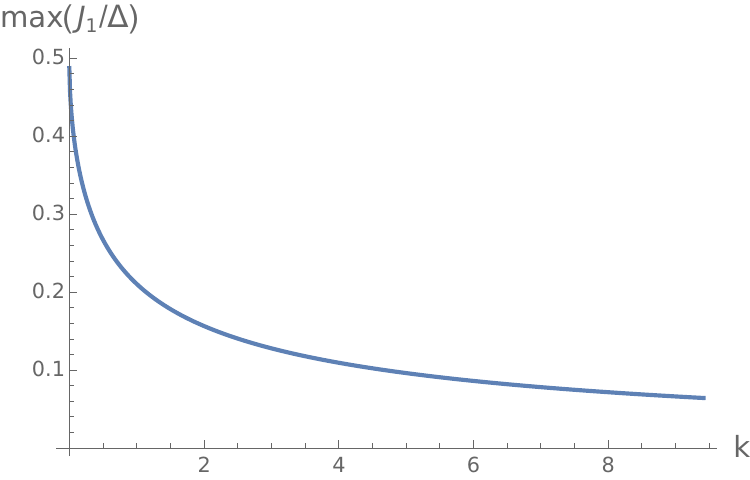}};
		\end{tikzpicture}
	}
	\caption{Left: The solid curves solve (\ref{eq:D3NS5-curves}) and host BPS geodesics, with $k\in\lbrace\frac{\pi}{100},\frac{\pi}{40},\frac{\pi}{10},\frac{\pi}{3}\rbrace$ increasing from the outer to the inner curves. The dashed curves in corresponding color solve (\ref{eq:geod-gen}) and host non-BPS geodesics. The black dashed lines are boundaries of $\Sigma$. Center: $J_1/\Delta$ along the curves for BPS geodesics in matching color; $J_2/\Delta=1-J_1/\Delta$. 
	Right: The maximal $J_1/\Delta$ for BPS geodesics as function of $k$.
	}
\end{figure}

For this solution there is a single regular point on the boundary of $\Sigma$ which hosts a single-charge BPS geodesic emerging from (\ref{eq:nice-cond-1}), (\ref{eq:nice-cond-2}), namely, with $k$ defined in (\ref{eq:D3NS5-BPS})
\begin{align}\label{eq:D3NS5-endpoints}
	z_1&=\ln\sqrt{1+\frac{2}{k}}~,
	&
	\tilde f_4^2&={\alpha'}N_5\left(\sqrt{k (  k+2)}+\cosh ^{-1}(k+1)\right).
\end{align}
The geodesic has angular momentum on the $SO(3)$ which acts along the NS5-directions transverse to the D3 branes, and is neutral under the $SO(3)$ acting transverse to D3 and NS5 branes.
The spectrum of nearby operators described by the pp-wave limit, determined by (\ref{eq:string-spectrum}), is given in (\ref{eq:D3NS5-spectrum}).

The more general condition for two-charge BPS geodesics in (\ref{eq:h1h2-cond-gen}) determines a single curve, 
\begin{align}\label{eq:D3NS5-BPS}
	\coth z+\coth \bar z-2&=2 k~,
	&
	k&\equiv \frac{\pi K}{N_5} e^{2\phi_0}=\frac{\lambda_{4d}}{4N_5^2}~.
\end{align}
It connects the regular boundary point (\ref{eq:D3NS5-endpoints}) hosting a single-charge BPS geodesic to the NS5-brane source at $z=0$, where the geometry exhibits the singular behavior appropriate for an NS5 brane. 
Plots are shown in fig.~\ref{fig:D3NS5curves}.
Each point along the curve hosts a null geodesic which satisfies the BPS condition (\ref{eq:BPS-ab}).
An explicit solution with $z=x+iy$ is
\begin{align}\label{eq:D3NS5-curves}
	y&=\frac{1}{2} \cos^{-1} \left(\cosh(2x)-\frac{\sinh(2x)}{k+1}\right)~,
	&
	0\leq x\leq\frac{1}{2}\ln\left(1+\frac{2}{k}\right)~.
\end{align}
Since the curves start and end on the same boundary of $\Sigma$, they interpolate between two points with $J_1=0$.
This is shown in fig.~\ref{fig:D3NS5-J12}. Along the curve there is a maximal $J_1/J$. 
The maximum as function of $k$ is shown in fig.~\ref{fig:D3NS5-J1max} and can be expressed as
\begin{align}\label{eq:D3NS5-J1-BPS-max}
	\left(\frac{J_1^2}{\Delta^2}\right)_{\rm max}&=\max_{\sqrt{k(k+1)}<x<k+1}\left[
	\frac{k(k+1)-x^2}{k (k+2)-x^2}\left(1-\frac{1}{2x}\ln\frac{k+1-x}{k+1+x}\right)
	\right].
\end{align}
It approaches $1/2$ for $k\rightarrow 0$ and zero for $k\rightarrow \infty$. For $J_1$ below this maximal value there are two BPS null geodesics.
For large $k$ the end point $z_1$ moves towards $z=0$ and the curve (\ref{eq:D3NS5-curves}) collapses.

\begin{figure}
	\subfigure[][]{\label{fig:D3NS5-J12-nonBPS}
		\includegraphics[width=0.4\linewidth]{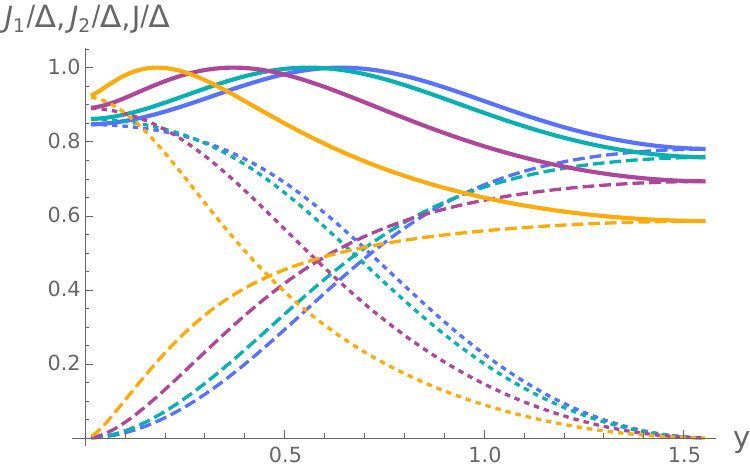}
	}
	\hskip 10mm
	\subfigure[][]{\label{fig:D3NS5-J-non-BPS}
		\includegraphics[width=0.36\linewidth]{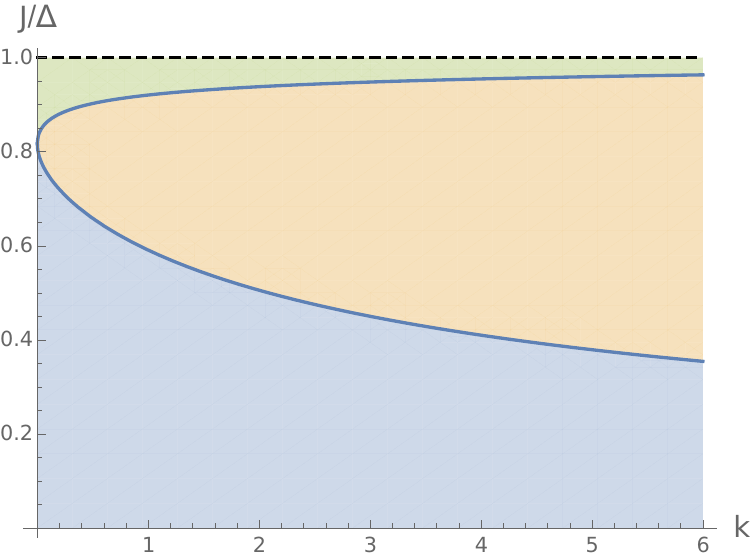}
		\put(-110,20){\footnotesize 0}
		\put(-110,80){\footnotesize 1}
		\put(-155,104){\footnotesize 2}
	}
	\caption{
	Left: $J_1/\Delta$ (dashed) and $J_2/\Delta$ (dotted) along the curves hosting non-BPS geodesics in fig.~\ref{fig:D3NS5curves}, in matching color. $(J_1+J_2)/\Delta$ (solid) reaches one where the curves for non-BPS and BPS geodesics intersect in fig.~\ref{fig:D3NS5curves}.	
	Right: Number of BPS geodesics for given $k$ and $J/\Delta$. The solid curves are (\ref{eq:D3NS5-single-charge-nonBPS}), (\ref{eq:D3NS5-single-charge-nonBPS-2}).\label{fig:D3NS5-nonBPS}
	}
\end{figure}

There are additional non-BPS geodesics resulting from (\ref{eq:geod-gen}), (\ref{eq:J12-non-BPS}). The solutions of (\ref{eq:geod-gen}) take the form of a single curve for each $k$ which connects the two boundary components of $\Sigma$, as shown in fig.~\ref{fig:D3NS5curves}. We find one single-charge geodesic for each $SO(3)$:
\begin{align}\label{eq:D3NS5-single-charge-nonBPS}
	\tilde z_1&=\frac{1}{4}\ln\left(1+\frac{2}{k}\right)+\frac{i\pi}{2}~, 
	& \frac{J_1^2}{\Delta^2}&=\frac{2}{3}-\frac{4}{9\sqrt{1+\frac{2}{k}}-3}~, & J_2&=0~,
\end{align}
and
\begin{align}\label{eq:D3NS5-single-charge-nonBPS-2}
	\tilde z_2&=\ln u~, & \frac{J_2^2}{\Delta^2}&=\frac{4u^2}{3ku^4-2(k-3)u^2-k-2}~, & J_1&=0~,
\end{align}
where $u>1$ is the solution to
\begin{align}\label{eq:D3NS5-single-charge-nonBPS-2a}
	u\,\frac{ 2-k \left(k u^4-k+3 u^2-3\right)}{\left(u^2-1\right) \left(3 k u^2+k+2\right)}&=\coth^{-1}u~.
\end{align}
The values of $J_i/\Delta$ realized by geodesics along the dashed curves in fig.~\ref{fig:D3NS5curves} are shown in fig.~\ref{fig:D3NS5-nonBPS}. We get a family of geodesics whose angular momenta interpolate between the ones given above.

In summary, we find one single-charge BPS geodesic for the $SO(3)$ rotating the NS5, whose spectrum of `nearby' operators is given by (\ref{eq:D3NS5-spectrum}), as well as one non-BPS single-charge geodesic for each $SO(3)$, with $J_i/\Delta$ given in (\ref{eq:D3NS5-single-charge-nonBPS}), (\ref{eq:D3NS5-single-charge-nonBPS-2}). We in addition found a family of two-charge BPS operators, which hint at further single-charge BPS operators at the NS5 source, and a family of non-BPS two-charge operators which interpolate between the two single-charge non-BPS operators.

\begin{figure}
	\includegraphics[width=0.33\linewidth]{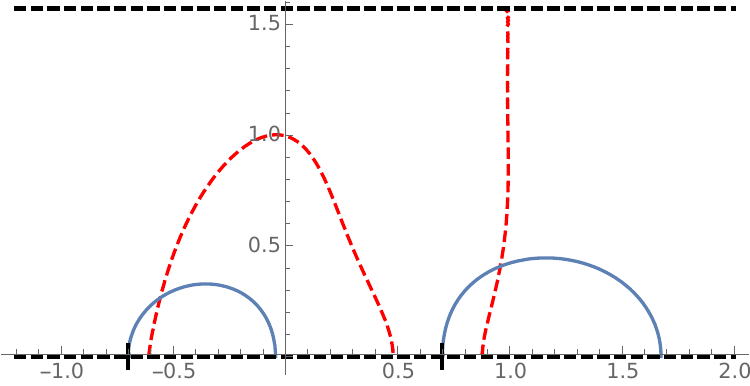}
	\hskip 20mm
	\includegraphics[width=0.33\linewidth]{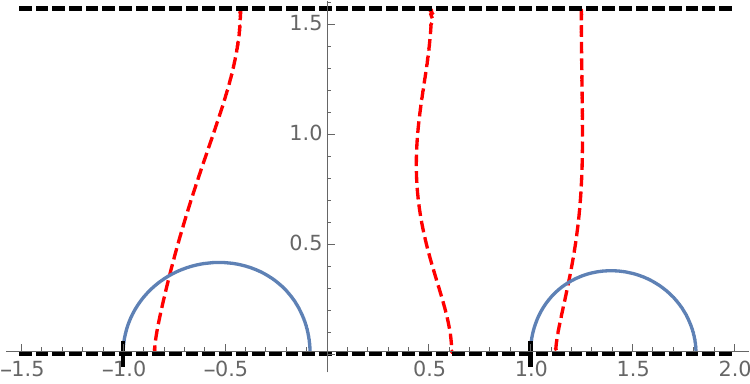}
	\put(-140,60){$\Sigma$}
	\put(-360,60){$\Sigma$}
	\caption{Curves hosting BPS geodesics (solid) and non-BPS geodesics (dashed) for $P=2$, $N_5^{(1)}=N_5^{(2)}=N_5/2$, $k=0.1$ and $\delta_1=-\delta_2=0.7$ (left), $\delta_1=-\delta_2=1$ (right). The curves hosting BPS geodesics connect the minima of $h_2$ to 5-brane sources, while the form of the dashed curves depends on the parameters.\label{fig:D3-NS5-P-plot}}
\end{figure}

\bigskip
\textbf{D3/NS5$^P$ BCFT:}
The discussion can be generalized to D3-branes ending on $P$ groups of NS5-branes with $N_5^{(p)}$ NS5-branes in each, such that within each group each NS5 has the same number of D3-branes ending on it. The supergravity duals are specified by
\begin{align}\label{eq:h1h2-D3NS5-BCFT-p}
	h_1&=-\frac{i\pi \alpha^\prime}{4} K e^z+\rm{c.c.}
	&
	h_2&=\frac{\pi\alpha^\prime}{4}K e^{z+2\phi_0}-\frac{\alpha^\prime}{4}\sum_{p=1}^P N_5^{(p)}\ln\tanh\left(\frac{z-\delta_p}{2}\right)+\rm{c.c.}
\end{align}
The locations of the NS5-branes, $z=\delta_p$, determine the numbers of D3-branes ending on each NS5 within the groups.
We illustrate the general features in fig.~\ref{fig:D3-NS5-P-plot}. In line with the discussion in sec.~\ref{sec:geod-Penrose}, we find $P$ single-charge BPS null geodesics on the boundary with the NS5 sources.

\bigskip
\textbf{D3/D5 BCFT:}
The D3/D5 BCFT is obtained by trading the NS5 branes in fig.~\ref{fig:D3NS5-brane-sol} for D5 branes.
The field theory will be discussed in sec.~\ref{sec:D3D5-ops}.
The supergravity solution is given by\footnote{The D3/D5 BCFT is S-dual to the D3/NS5 BCFT. In supergravity S-duality swaps $h_1\leftrightarrow h_2$ with $z\rightarrow \frac{i\pi}{2}+\bar z$. The functions (\ref{eq:h1h2-D3D5-BCFT}) are related to (\ref{eq:h1h2-D3NS5-BCFT}) by S-duality combined with $\phi_0\rightarrow -\phi_0$ to maintain the meaning of $\phi_0$.}
\begin{align}\label{eq:h1h2-D3D5-BCFT}
	h_1&=-\frac{i\pi \alpha'}{4}K e^{z-2\phi_0}-\frac{\alpha'}{4}N_5\ln\tanh\left(\frac{i\pi}{4}-z\right)+\rm{c.c.}
	&
	h_2&=\frac{\pi \alpha'}{4}K e^z+\rm{c.c.}
\end{align}
The asymptotic value of the dilaton and the 4d $\mathcal N=4$ SYM `t Hooft coupling are given by
\begin{align}
	\lim_{\Re(z)\rightarrow\infty}e^{2\phi}&=e^{2\phi_0}~,
	&
	\lambda_{4d}&=4\pi N_5K e^{2\phi_0}~.
\end{align}

The condition for BPS geodesics in (\ref{eq:h1h2-cond-gen}) leads to curves which are obtained from those in fig.~\ref{fig:D3NS5curves} by a vertical reflection on the strip, namely
\begin{align}\label{eq:D3D5-geod-curve}
	\tanh z+\tanh \bar z-2&=2k~,
	&
	k&\equiv \frac{\pi K}{N_5} e^{-2\phi_0}=\frac{4\pi^2 K^2}{\lambda_{4d}}~.
\end{align}
The definition of $k$ has changed and the roles of $J_1$ and $J_2$ are exchanged: The curves for BPS geodesics have $J_2=0$ at both ends and the plots in fig.~\ref{fig:D3NS5-J12}, \ref{fig:D3NS5-J1max} carry over to D3/D5 with $J_1\leftrightarrow J_2$ and the modified $k$.
For the single-charge BPS geodesic we find
\begin{align}\label{eq:D3D5-single-charge}
	z_1&=\ln\sqrt{1+\frac{2}{k}}+\frac{i\pi}{2}~,
	&
	\tilde f_4^4&={\alpha'}^2N_5^2k e^{4\phi_0} \left(1+\frac{\cosh ^{-1}( k+1)}{\sqrt{k(k+2)}}\right).
\end{align}
This leads, via (\ref{eq:string-spectrum}) to (\ref{eq:D3D5-spectrum}).
The spectrum of null geodesics for D3/D5 is generally obtained from the D3/NS5 results by replacing $k$ with the definition in (\ref{eq:D3D5-geod-curve}) and exchanging the two $SO(3)$'s.
 The relation generalizes to D3/D5$^P$. It may be interesting to compare the BPS results to \cite{Hatsuda:2024uwt}.

\subsection{D3/D5/NS5 BCFT}\label{sec:D3D5NS5-sugra}

\begin{figure}
	\subfigure[][]{\label{fig:D3D5NS5-sol}
		\begin{tikzpicture}[scale=1]
			\shade [right color=3dcolor!100,left color=3dcolor!100] (-0.3,0)  rectangle (0.3,-2);
			
			\shade [ left color=3dcolor! 100, right color=4dcolor! 100] (0.3-0.01,0)  rectangle (2,-2);
			\shade [ right color=3dcolor! 100, left color=4dcolor! 100] (-0.3+0.01,0)  rectangle (-2,-2);

			\draw[thick] (-2,0) -- (2,0);
			\draw[thick] (-2,-2) -- (2,-2);
			\draw[dashed] (2,-2) -- +(0,2);
			\draw[thick] (-2,-2) -- +(0,2);
			
			\node at (-0.5,-0.5) {$\Sigma$};
			\node at (2.5,-0.65) {\footnotesize $AdS_5$};
			\node at (2.5,-1) {\footnotesize $\times$};
			\node at (2.5,-1.35) {\footnotesize $S^5$};			
			
			\draw[very thick] (0,-0.08) -- (0,0.08) node [anchor=south] {\footnotesize D5};
			\draw[thick] (0,-1.92) -- (0,-2.08) node [anchor=north] {\footnotesize NS5};
		\end{tikzpicture}
	}\hskip 20mm
	\subfigure[][]{\label{fig:D3D5NS5-brane}
		\begin{tikzpicture}[y={(0cm,1cm)}, x={(0.707cm,0.707cm)}, z={(1cm,0cm)}, scale=1.1]
			\draw[gray,fill=gray!100,rotate around={-45:(0,0,1.8)}] (0,0,1.8) ellipse (1.8pt and 3.5pt);
			\draw[gray,fill=gray!100] (0,0,0) ellipse (1.5pt and 3pt);
			
			\foreach \i in {-0.05,0,0.05}{ \draw[thick] (0,-1,\i) -- (0,1,\i);}

			\foreach \i in {-0.075,-0.025,0.025,0.075}{ \draw (-1.1,\i,1.8) -- (1.1,\i,1.8);}
			
			\foreach \i in  {-0.075,-0.045,-0.015,0.015,0.045,0.075}{ \draw (0,1.4*\i,0) -- (0,1.4*\i,1.8+\i);}
			
			\foreach \i in {-0.045,-0.015,0.015,0.045}{ \draw (0,1.4*\i,1.8+\i) -- (0,1.4*\i,4);}

			\node at (-0.18,-0.18,3.4) {\scriptsize $N_{\rm D3}^\infty$ D3};
			\node at (1.0,0.3,1.8) {\scriptsize $N_5$ D5};
			\node at (0,-1.25) {\footnotesize $N_5$ NS5};
			\node at (0.18,0.18,0.8) {{\scriptsize $N_{\rm D3}^0$ D3}};
		\end{tikzpicture}
	}
	\caption{D3/D5/NS5 BCFT: schematic form of supergravity solution (left) and brane construction (right). $N_{\rm D3}^\infty=2N_5K$ semi-infinite D3-branes end on $N_5$ D5-branes and $N_5$ NS5-branes, with $N_{\rm D3}^0=N_5K+N_5^2/2$.}
\end{figure}
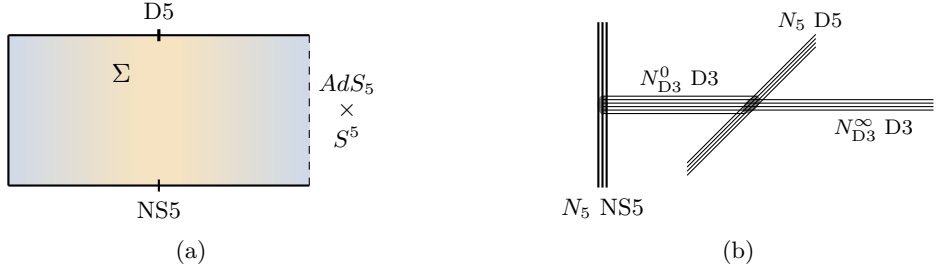

The D3/D5/NS5 BCFT is engineered by $2N_5K$ semi-infinite D3-branes ending on a combination of $N_5$ D5-branes and $N_5$ NS5-branes, fig.~\ref{fig:D3D5NS5-brane}.
Suspended between the D5 and NS5 branes are $N_5K+N_5^2/2$ D3-branes. The BCFT emerges in limit where the 5-branes become coincident in the horizontal direction and the field theory is discussed in sec.~\ref{sec:D3D5NS5-ops}.
The harmonic functions are 
\begin{align}\label{eq:h1h2-BCFT}
	h_1&=-\frac{i\pi\alpha^\prime}{4}K e^z-\frac{\alpha^\prime}{4}N_5\ln\tanh\left(\frac{i\pi}{4}-\frac{z}{2}\right)+\rm{c.c.}
	\nonumber\\
	h_2&=\frac{\pi \alpha^\prime}{4} K e^z-\frac{\alpha^\prime}{4}N_5\ln\tanh\left(\frac{z}{2}\right)+\rm{c.c.}
\end{align}
The solution is illustrated in fig.~\ref{fig:D3D5NS5-sol}.
This setup was used in \cite{Uhlemann:2021nhu} to analyze black hole information transfer. 
We set the 4d $\mathcal N=4$ SYM coupling to one for simplicity; with this choice the setup is self-S-dual. The 4d coupling can be changed using $SL(2,\ZZ)$ (see discussion in \cite{Karch:2022rvr}).

\begin{figure}
	\subfigure[][]{\label{fig:D3D5NS5-curves}
		\begin{tikzpicture}
			\node at (0,0) {\includegraphics[width=0.36\linewidth]{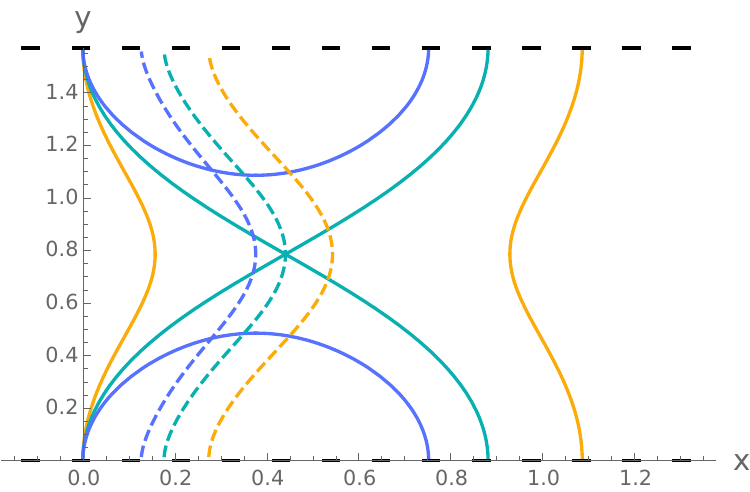}};
			\node at (2,1) {$\Sigma$};
		\end{tikzpicture}
	}
	\hskip -2mm
	\subfigure[][]{\label{fig:D3D5NS5-J1J2-3}
		\includegraphics[width=0.23\linewidth]{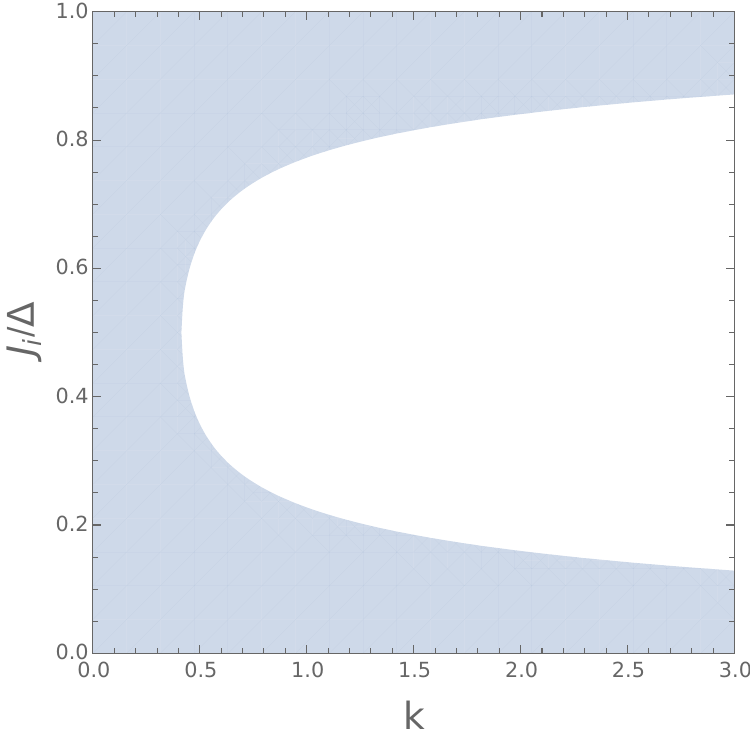}
	}
	\hskip 0mm
	\subfigure[][]{\label{fig:D3D5NS5-J1J2-nonBPS-single}
		\includegraphics[width=0.34\linewidth]{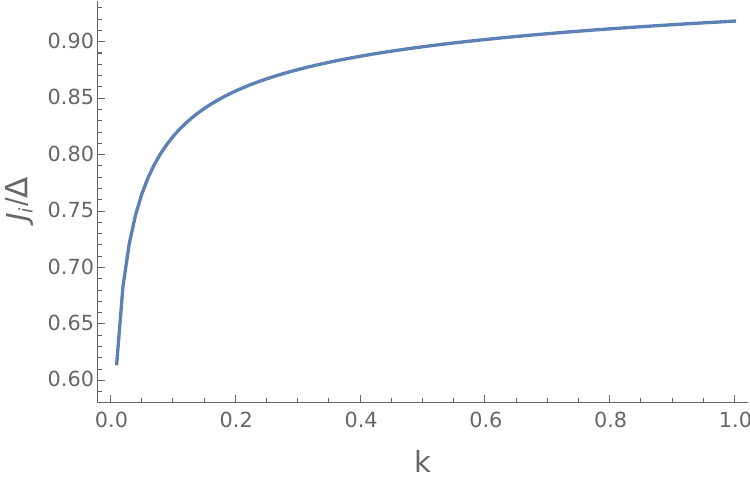}
	}
	\caption{Left: The solid curves host BPS geodesics for $k=k_\star+\lbrace -\frac{\pi}{20},0, \frac{\pi}{20}\rbrace$ with $k_\star$ in (\ref{eq:k-crit-D5NS5D3}), with the color indicating $k$.
	For $k<k_\star$ (yellow) one curve connects the 5-brane sources at $x=0$, another connects regular points on opposite boundaries. 
	For $k>k_\star$ (blue), the two curves connect points on the same boundaries.
	The dashed curves host non-BPS geodesics.
	Center: For $(k,J_i/\Delta)$ in the blue-shaded region there is a pair of BPS geodesics, in the white region there are none. 
	Right: $J_i/\Delta$ for single-charge non-BPS geodesics.}
\end{figure}

The points on $\Sigma$ hosting BPS null geodesics, determined from (\ref{eq:h1h2-cond-gen}), are curves satisfying
\begin{align}\label{eq:D3D5NS5-bps-cond}
	 k^2 \left|e^{4 z}-1\right|^2-2 k \left(e^{4 z}+e^{4\bar z}-2\right)-4 e^{2 (z+\bar z)}+4&=0~,
	&
	k\equiv \frac{\pi K}{N_5}~.
\end{align}
The solutions, shown in fig.~\ref{fig:D3D5NS5-curves}, are symmetric under vertical reflection of $\Sigma$.
For each $k$ there are two regular points on the boundaries of $\Sigma$ hosting single-charge BPS geodesics, given by
\begin{align}\label{eq:D3D5NS5-nice}
	z_1&=\ln\sqrt{1+\frac{2}{k}}~, 
	& 
	\tilde f_4^2&={\alpha'}N_5\left(\sqrt{k (k+2)}+\cosh^{-1}(k+1)\right)~,
	\nonumber\\
	z_2&=z_1+\frac{i\pi}{2}~,
	&
	\tilde f_4^4&={\alpha'}^2N_5^2k\frac{(k+2)^2}{(k+1)^2}\left(1+\frac{\cosh ^{-1}(k+1)}{\sqrt{k (k+2)}}\right)~.
\end{align}
The spectrum of operators described by the associated pp-waves is determined by (\ref{eq:string-spectrum}), leading to (\ref{eq:D3D5NS5-spectrum}).
We again find the 5-brane sources at $z=0,\frac{i\pi}{2}$ as solutions to (\ref{eq:D3D5NS5-bps-cond}), hinting at additional single-charge BPS geodesics not covered by the discussion of Penrose limits in sec.~\ref{sec:geod-Penrose}.

How the points in (\ref{eq:D3D5NS5-nice}) and the 5-brane sources are connected through $\Sigma$ by curves satisfying (\ref{eq:D3D5NS5-bps-cond}) depends on $k$.
For small $k$ there is one curve connecting $z_1$ to $z_2$ and another connecting the 5-brane sources.
For large $k$, one curve connects $z_1$ to the NS5 source while the other connects $z_2$ to the D5 source. This is shown in fig.~\ref{fig:D3D5NS5-curves}.
The critical value separating the two regimes is
\begin{align}\label{eq:k-crit-D5NS5D3}
	k_\star\equiv \Big(\frac{\pi K}{N_5}\Big)_\star&=\sqrt{2}-1~.
\end{align}
For $k<k_\star$ the curves connect points on opposite boundaries of $\Sigma$, so all combinations of $J_{1/2}/\Delta$ are realized, similar to the $\rm AdS_5\times S^5$ case. For $k>k_\star$ the curves connect points on the same boundary, so only certain $J_i/J$ are realized, similar to the D3/NS5 case. This is shown in fig.~\ref{fig:D3D5NS5-J1J2-3}.

There are also non-BPS geodesics resulting from (\ref{eq:geod-gen}). The solutions of (\ref{eq:geod-gen}) comprise a single curve for each $k$ which connects the two boundary components of $\Sigma$ (fig.~\ref{fig:D3D5NS5-curves}).
This identifies two single-charge non-BPS geodesics connected by a one-parameter family of two-charge non-BPS geodesics.
The curves are symmetric under a vertical reflection on $\Sigma$ as required by S-duality.
For $k>k_\star$ the curve hosting non-BPS geodesics intersects the curves hosting BPS geodesics, for $k<k_\star$ the curves do not intersect. Plots of $J_i/\Delta$ for the single-charge non-BPS geodesics are in fig.~\ref{fig:D3D5NS5-J1J2-nonBPS-single}.

\textbf{3d SCFT limit:}
A special case arises for $K=0$. The semi-infinite D3-branes in fig.~\ref{fig:D3D5NS5-brane} disappear, leaving only D3-branes suspended between D5 and NS5 branes engineering a 3d $T_\rho^\sigma[SU(N)]$ SCFT.
The boundary points in (\ref{eq:D3D5NS5-nice}) move to infinity for $K\rightarrow 0$ and disappear for $K=0$, as does the curve connecting them. Likewise, the curve hosting non-BPS geodesics disappears. 
Only $x=0$ remains as solution to (\ref{eq:h1h2-cond-gen}), and both end points on the boundaries are at 5-brane sources.

\subsection{\texorpdfstring{D5$^2$/NS5$^2$}{D52/NS52} interface}\label{sec:D52-NS52-sugra}

\begin{figure}
	\subfigure[][]{\label{fig:D52NS52-sol}
		\begin{tikzpicture}[scale=1]
			\shade [right color=3dcolor!100,left color=3dcolor!100] (-0.3,0)  rectangle (0.3,-2);
			
			\shade [ left color=3dcolor! 100, right color=4dcolor! 100] (0.3-0.01,0)  rectangle (2,-2);
			\shade [ right color=3dcolor! 100, left color=4dcolor! 100] (-0.3+0.01,0)  rectangle (-2,-2);
			
			\draw[thick] (-2,0) -- (2,0);
			\draw[thick] (-2,-2) -- (2,-2);
			\draw[dashed] (2,-2) -- +(0,2);
			\draw[dashed] (-2,-2) -- +(0,2);
			
			\node at (-0.5,-0.5) {$\Sigma$};
			\node at (2.5,-0.65) {\footnotesize $AdS_5$};
			\node at (2.5,-1) {\footnotesize $\times$};
			\node at (2.5,-1.35) {\footnotesize $S^5$};
			
			\node at (-2.5,-0.65) {\footnotesize $AdS_5$};
			\node at (-2.5,-1) {\footnotesize $\times$};
			\node at (-2.5,-1.35) {\footnotesize $S^5$};

			\draw[very thick] (-0.7,-0.08) -- (-0.7,0.08) node [anchor=south] {\footnotesize NS5};
			\draw[very thick] (0.7,-0.08) -- (0.7,0.08) node [anchor=south] {\footnotesize NS5};
			\draw[thick] (-0.7,-1.92) -- (-0.7,-2.08) node [anchor=north] {\footnotesize D5};
			\draw[thick] (0.7,-1.92) -- (0.7,-2.08) node [anchor=north] {\footnotesize D5};
		\end{tikzpicture}
	}\hskip 15mm
	\subfigure[][]{\label{fig:D52NS52-brane}
		\begin{tikzpicture}[y={(0cm,1cm)}, x={(0.707cm,0.707cm)}, z={(1cm,0cm)}, xscale=1,yscale=1.1]
			\draw[gray,fill=gray!100] (0,0,-0.5) ellipse (1.8pt and 3pt);
			\draw[gray,fill=gray!100] (0,0,1) ellipse (1.8pt and 4.5pt);
			\draw[gray,fill=gray!100,rotate around={-45:(0,0,2.5)}] (0,0,2.5) ellipse (1.8pt and 5pt);
			\draw[gray,fill=gray!100,rotate around={-45:(0,0,4)}] (0,0,4) ellipse (1.8pt and 3pt);				
			
			\foreach \i in {-0.05,0,0.05}{ \draw[thick] (0,-1,-0.5+\i) -- (0,1,-0.5+\i);}
			\foreach \i in {-0.05,0,0.05}{ \draw[thick] (0,-1,1+\i) -- (0,1,1+\i);}

			\foreach \i in {-0.075,-0.025,0.025,0.075}{ \draw (-1.1,\i,2.5) -- (1.1,\i,2.5);}
			\foreach \i in {-0.075,-0.025,0.025,0.075}{ \draw (-1.1,\i,4) -- (1.1,\i,4);}
			
			\foreach \i in {-0.06,-0.03,0,0.03,0.06}{ \draw (0,1.4*\i,-0.5) -- (0,1.4*\i,1);}
			\foreach \i in {-0.1,-0.075,-0.045,-0.015,0.015,0.045,0.075,0.1}{ \draw (0,1.4*\i,1) -- (0,1.4*\i,2.5+\i);}
			\foreach \i in {-0.06,-0.03,0,0.03,0.06}{ \draw (0,1.4*\i,2.5) -- (0,1.4*\i,4);}
			
			\foreach \i in {-0.03,0,0.03}{ \draw (0,1.4*\i,4) -- (0,1.4*\i,5.5);}
			\foreach \i in {-0.03,0,0.03}{ \draw (0,1.4*\i,-0.5) -- (0,1.4*\i,-2);}
			
			\node at (0,-1.25,-0.5) {\scriptsize $N_5/2$ NS5};
			\node at (0,-1.25,1) {\scriptsize $N_5/2$ NS5};
			\node at (1.0,0.35,2.5) {\scriptsize  $N_5/2$ D5};
			\node at (1.0,0.35,4) {\scriptsize  $N_5/2$ D5};
			\node at (0.2,0.2,1.75) {{\scriptsize $N_{\rm D3}^0$}};
			\node at (0,0.28,0.25) {{\scriptsize $N_{\rm D3}^1$}};
			\node at (0,0.28,3.5) {{\scriptsize $N_{\rm D3}^1$}};
			
			\node at (0,0.28,5) {{\scriptsize $N_{\rm D3}^\infty$}};
			\node at (0,0.28,-1.25) {{\scriptsize $N_{\rm D3}^\infty$}};
		\end{tikzpicture}
	}
	\caption{Left: supergravity solutions (\ref{eq:D52NS52-h12}).
		Right: Brane configuration with D3-branes intersecting two groups of D5-branes and two groups of NS5-branes with D3-branes suspended between them.}
\end{figure}
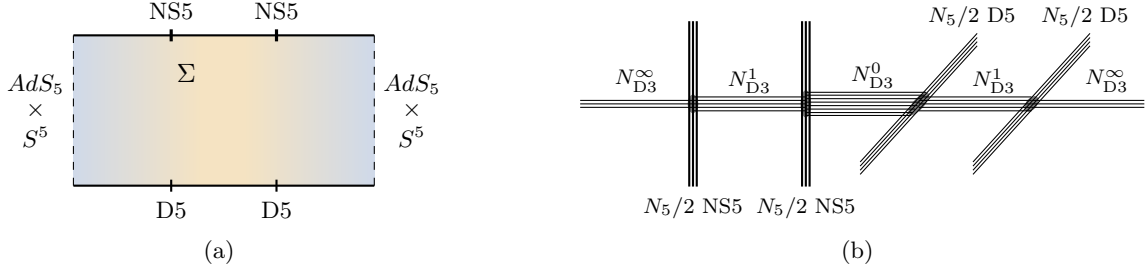

The last example is an ICFT engineered by D3-branes intersecting two groups of D5-branes and two groups of NS5-branes, as shown in fig.~\ref{fig:D52NS52-brane}. 
The numbers of D3-branes ending on each 5-brane are identical within each group, but differ between the two groups of D5 and NS5 branes.
This leads to an interface with 3d degrees of freedom. 
The interface EE was studied in \cite{Uhlemann:2023oea}.

For simplicity we set the 4d couplings to the self-dual values on both sides; arbitrary couplings on either side can be realized along the same lines.
The harmonic functions are
\begin{align}\label{eq:D52NS52-h12}
	h_1&=\frac{\pi\alpha'}{2}K\cosh z-\frac{\alpha'}{4}\frac{N_5}{2}\ln\left[\tanh \left(\frac{z-\delta}{2}\right)\tanh \left(\frac{z+\delta}{2}\right)\right]+\mathrm{c.c.}
	\nonumber\\
	h_2&=-\frac{i\pi\alpha'}{2}K\sinh z-\frac{\alpha'}{4}\frac{N_5}{2}\ln\left[\tanh \left(\frac{i\pi}{4}-\frac{z-\delta}{2}\right)\tanh \left(\frac{i\pi}{4}-\frac{z+\delta}{2}\right)\right]+\mathrm{c.c.}
\end{align}
The solutions, with pairs of 5-brane sources at $z=\pm\delta$ and $z=\pm \delta+\frac{i\pi}{2}$, are illustrated in fig.~\ref{fig:D52NS52-sol}.
The 10d metric is invariant under $x\rightarrow -x$, which realizes a reflection across the interface in the field theory, and under $y\rightarrow \frac{\pi}{2}-y$.
The D3-brane numbers obtained via \cite[(4.27)]{Assel:2011xz} are
\begin{align}\label{eq:D52NS52-ND3}
	N_{\rm D3}^\infty&=\pi K^2+2N_5 K \cosh \delta~, &
	N_{\rm D3}^0&=N_{\rm D3}^\infty+\frac{N_5^2}{2}~,& 
	N_{\rm D3}^1&=N_{\rm D3}^\infty+N_5^2\frac{\Delta_{k,\delta}}{4}~, 
\end{align}
where we defined $\Delta_{k,\delta}=\frac{1}{2}+\frac{2}{\pi}\arctan e^{-2\delta}-4k\sinh\delta$ with $k\equiv K/N_5$.
For $K=0$ the ICFT reduces to a 3d $T_\rho^\sigma[SU(N)]$ SCFT; this 3d SCFT was used for wedge holography in \cite{Uhlemann:2021nhu}.

The general conditions for single-charge BPS geodesics, (\ref{eq:nice-cond-1}) and (\ref{eq:nice-cond-2}), yield 3 pairs of single-charge BPS geodesics with one geodesic for each $SO(3)$. The first pair,
\begin{align}\label{eq:D52-NS52-interface-f4t-1a}
	z_1&=0~,& 	\tilde f_4^2&=\alpha' N_5\left(2\pi k-\ln\tanh\frac{\delta}{2}\right)~,
	& k&\equiv\frac{K}{N_5}~,
	\\
	z_2&=\frac{i\pi}{2}~, & \tilde f_4^4&=
	\frac{{\alpha'}^2N_5^2(1+d)^2\tanh ^2\delta }{\cosh\delta+2 \pi  k \sinh ^2\delta}
	\left(2 \pi  k-\ln \tanh
	\frac{\delta}{2}\right),
	&
	d&\equiv 2\pi k \cosh\delta~,
	\label{eq:D52-NS52-interface-f4t-1b}
\end{align}
is invariant under $x\rightarrow -x$. The remaining points are
\begin{align}\label{eq:D52-NS52-interface-f4t-2a}
	z_{3/4}&=\pm\cosh^{-1}\frac{\sqrt{d(1+d)}}{2\pi k}~,
	&
	\tilde f_4^2&=\alpha'N_5\left(\sqrt{d\left(1+d\right)}+\sinh^{-1}\sqrt{d}\right)~,
\end{align}
and
\begin{align}\label{eq:D52-NS52-interface-f4t-2b}
	z_{5/6}&=z_{3/4}+\frac{i\pi}{2}~, &
	\tilde f_4^4&=2{\alpha'}^2N_5^2d\frac{\left(d (d+1)-2 \pi ^2 k^2\right)^2 \left(d (d+1)+\sqrt{d (d+1)} \sinh ^{-1}\sqrt{d}\right)}{\left(d
		(d+1)-4 \pi ^2 k^2\right) \left(d (2 d+1)-4 \pi ^2 k^2\right)^2}
	~.
\end{align}
The reflection $x\rightarrow -x$ exchanges $z_{3,5}\leftrightarrow z_{4,6}$.
The pp-wave string spectrum is determined by (\ref{eq:string-spectrum}).
For $\delta\rightarrow 0$ the 5-brane sources on each boundary merge onto the geodesics in (\ref{eq:D52-NS52-interface-f4t-1a}), (\ref{eq:D52-NS52-interface-f4t-1b}).

\begin{figure}
	\centering
	\includegraphics[width=0.3\linewidth]{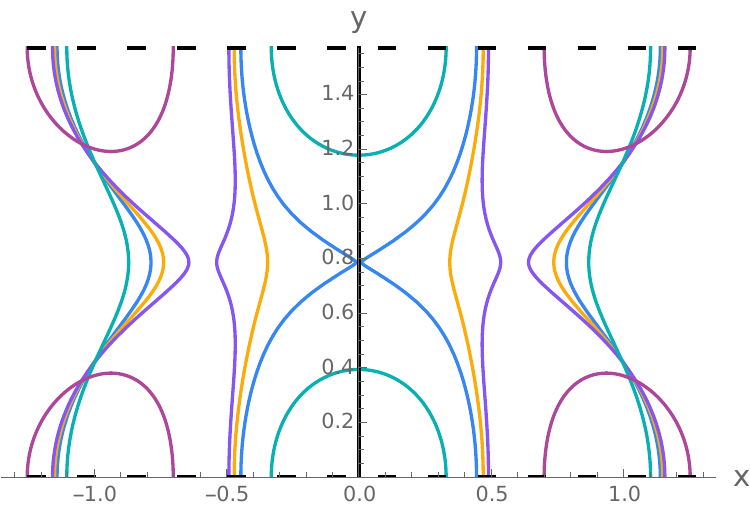}
	\hskip 3mm
	\includegraphics[width=0.3\linewidth]{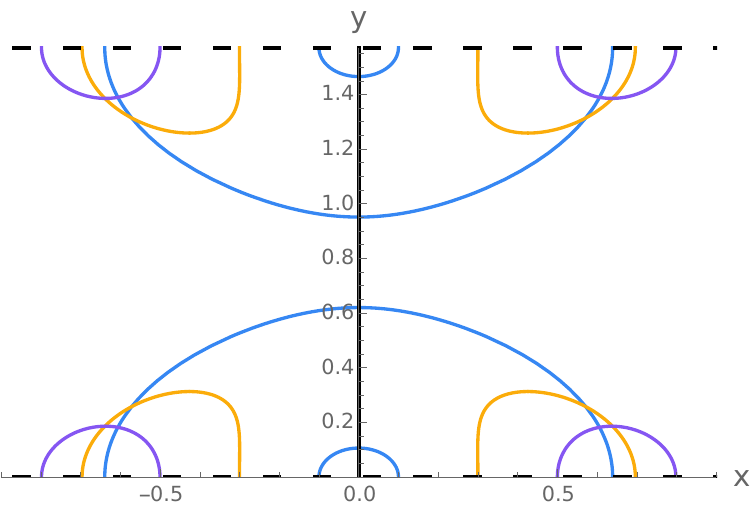}
	\hskip 3mm
	\includegraphics[width=0.3\linewidth]{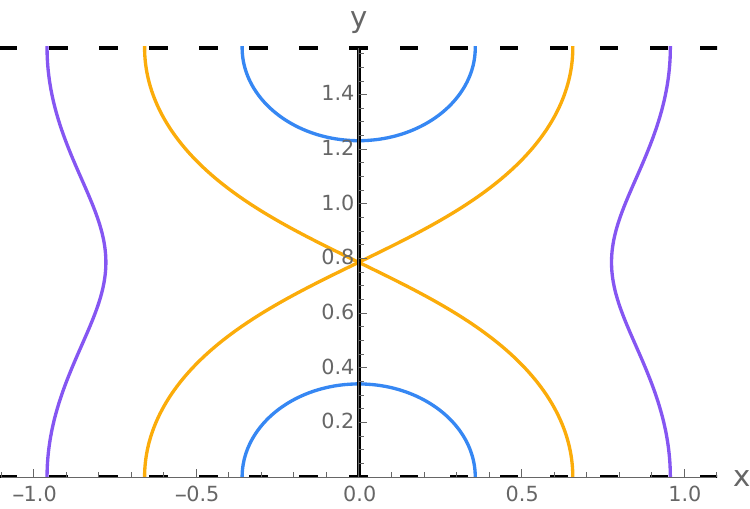}
	\caption{Curves hosting 2-charge BPS geodesics for D5$^2$/NS5$^2$ ICFTs with various $\delta$ and $k=0.1$ (left), $k=0.35$ (center), and for 3d SCFTs with $K=0$ (right).  $\Re(z)=0$ is always included.\label{fig:D52NS52-ICFT}}
\end{figure}

2-charge BPS geodesics can be placed along the entire line $z=iy$, which is fixed by the $x\rightarrow -x$ symmetry and a solution to (\ref{eq:h1h2-cond-gen}). 
Geodesics placed on this curve can realize all BPS combinations of $J_{1/2}$ charges.
The points $z_{3,4,5,6}$ and the 5-brane sources are connected by 4 additional curves which host 2-charge BPS geodesics. Depending on $k$ and $\delta$, the curves either connect points on the same boundary of $\Sigma$ or on opposite boundaries (fig.~\ref{fig:D52NS52-ICFT}).
The critical values are determined by
\begin{align}\label{eq:D52-NS52-k-delta-crit}
	\pi k &= \cosh (\delta ) \sech(2 \delta ) (2 \sech(2 \delta )-1)~.
\end{align}
This is where the number of solutions at $y=\frac{\pi}{4}$ changes. Additional non-BPS null geodesics arise from (\ref{eq:geod-gen}). They can be discussed along similar lines as for previous solutions.

For the 3d SCFT case $K=0$ only the single-charge BPS geodesics in (\ref{eq:D52-NS52-interface-f4t-1a}), (\ref{eq:D52-NS52-interface-f4t-1b}) remain. 
The points $z_{3,4,5,6}$ move to large $|\Re(z)|$ for $K\rightarrow 0$ and disappear at $K=0$.
In addition to $x=0$ there are two curves solving (\ref{eq:h1h2-cond-gen}) which connect pairs of 5-brane sources (fig.~\ref{fig:D52NS52-ICFT}).
For $\delta>\delta_\star$, with
\begin{align}\label{eq:delta-crit}
	\delta_\star&=\frac{1}{2}\ln(2+\sqrt{3})~,
\end{align}
the curves connect 5-brane sources on opposite boundaries of $\Sigma$ and the 2-charge geodesics realize all BPS combinations of $J_i/\Delta$. For $\delta<\delta_\star$ the curves connect points on the same boundary and there is a gap.

\section{4d \texorpdfstring{$\mathcal N=4$}{N=4} SYM B/I/dCFT}\label{sec:fieldtheory}

In this section we discuss concrete $\mathcal N=4$ SYM boundary, defect and interface CFTs, interpret the holographic results of sec.~\ref{sec:examples}, and discuss the field theory realization of the BMN-like sectors identified in sec.~\ref{sec:examples}. 
The analysis will not be exhaustive but we identify for each theory a set of candidate seed operators whose numbers and features naturally match the holographic discussion of BMN-like sectors.
We start with general preliminaries before turning to concrete field theories.

The boundaries and defects preserve 3d $\mathcal N=4$ (defect) superconformal symmetry with $\rm SO(3)_C\times SO(3)_H$ R-symmetry.
The vector multiplet of the 4d $\mathcal N=4$ SYM ambient CFT decomposes into a 3d vector multiplet, containing a vector field and an $SO(3)_C$ triplet of real scalars, and a 3d hypermultiplet, containing an $SO(3)_H$ doublet of complex scalars.
The scalars $\phi^I$ of 4d $\mathcal N=4$ SYM forming a vector of the $SO(6)$ R-symmetry split into two groups \cite[sec.~2]{DHoker:2007zhm}, \cite[sec.~2]{Gaiotto:2008sd},
\begin{align}\label{eq:N4SYM-scalar-split}
	\lbrace\phi^I, I=1,\ldots,6\rbrace 
	\quad 
	&\rightarrow 
	\quad
	\vec{X}=(X^1,X^2,X^3)
	\ \ \cup \ \
	\vec{Y}=(Y^1,Y^2,Y^3)~.
\end{align}
We split the 4d spacetime coordinates as $x^{\mu=0,1,2}$ and $x^3$, with the defect at $x^3=0$.
The 4d vector field $A_\mu$ is an $SO(6)$ singlet and decomposes accordingly.
The 4d vector multiplet decomposes as
\begin{align}\label{eq:4d-vector-split}
	(A_\mu,A_3,\vec{X},\vec{Y})+{\rm fermions}
	\quad &\rightarrow\quad
	(A_{0,1,2},\vec{Y})+{\rm fermions}
	\quad\cup\quad
	(A_3,\vec{X})+{\rm fermions}
\end{align}
The first set of fields forms a 3d vector multiplet, the second  a 3d hypermultiplet.
The scalars $\vec{X}$ describe fluctuations of the D3-branes in the D5 directions and are charged under $SO(3)_H\sim SO(3)_X$, while $\vec{Y}$ describe fluctuations in the NS5 directions charged under $SO(3)_C\sim SO(3)_Y$.

The pp-wave limits identified in sec.~\ref{sec:examples} describe states on the cylinder.
For CFTs with defects, these relate to operators on the defect. Consider a 4d CFT on the cylinder $\RR\times S^3$, with defect along $\RR\times S^2$. If the 4d theory on one half space is trivial this is a BCFT, if both are trivial a 3d CFT. 
The conformal transformation underlying the state-operator map, from the cylinder to the plane in polar coordinates, maps the defect to an $\RR^3$ in $\RR^4$, and past infinity to the origin of  $\RR^3$:
\begin{center}
	\begin{tikzpicture}[xscale=0.85,yscale=0.75]
		\draw[thick] (0,0) ellipse (1.0cm and 0.5cm);
		\draw[white,fill=white] (-1,0) rectangle (1,1);
		\draw[dashed, thick] (0,0) ellipse (1.0cm and 0.5cm);
		
		\draw[blue,very thick] (-0.05,1.5) -- (-0.05,-0.5);
		\draw[blue,very thick,dashed] (0.05,2.5) -- (0.05,0.5);
		
		\draw[thick] (0,2) ellipse (1.0cm and 0.5cm);
		\draw[thick] (-1,0) -- (-1,2);
		\draw[thick] (1,0) -- (1,2);
		
		\draw [->] (2.75,1) -- (3.5,1);
		
		\draw[dashed] (5,-0.25) rectangle (7.5,2.25);
		\draw[blue,very thick] (5,1) -- (7.5,1);
		\node at (6.25,1) {\sf X};
	\end{tikzpicture}
\end{center}
States on the cylinder prepared by a path integral with boundary conditions at past infinity then map to an operator insertion at the origin of the $\RR^3$ in $\RR^4$, i.e.\ on the defect.

BCFTs may generally contain operators on the boundary which arise as boundary limit of 4d ambient CFT operators, as well as genuine 3d operators which do not arise as limit of ambient CFT operators. 
They generally mix in the identification with closed string modes at the defect.\footnote{As discussed for D3/probe D5 defects in \cite[sec.~5]{DeWolfe:2001pq}, where identifying the graviton with the energy-momentum tensor and the dilaton with the Lagrangian leads to both picking up defect contributions.}
We expect seed operators corresponding to closed string pp-wave vacua of BMN-like sectors to generally be invariant under non-Abelian global symmetries, while operators transforming under non-Abelian global symmetries should correspond to open string modes at the 5-brane sources.

The general 3d $\mathcal N=4$ superconformal multiplets were classified in \cite{Dolan:2008vc,Cordova:2016emh}.
From \cite[table 6]{Cordova:2016emh}, we highlight the
short $B_1$ and $A_2$ multiplets with scalar primaries
denoted as
\begin{align}\label{eq:3dN4-multiplets}
	B_1[0]^{(R;R')}_{\Delta}:&\quad	\Delta=\frac{1}{2}(R+R^\prime)~,
	&
	A_2[0]^{(R;R')}_{\Delta}:&\quad	\Delta=\frac{1}{2}(R+R^\prime)+1~,
\end{align}
where $R,R'$ are the $SO(3)_H\times SO(3)_C$ $R$-charges in a convention with $R=1$ for an $SU(2)$ doublet. The relation to our convention with $J=1$ for the triplet is
\begin{align}
	R&=2J_1~, & R^\prime&=2J_2~.
\end{align}
Free hypermultiplets are $B_1[0]_{1/2}^{(1;0)}$, twisted hypermultiplets $B_1[0]_{1/2}^{(0;1)}$.
$B_1$ multiplets with $R,R'\geq 2$ and $A_2$ multiplets appear in the decomposition of long multiplets hitting BPS bounds \cite[(2.14)]{Cordova:2016emh}.
$B_1$ multiplets with  $R<1$ or $R'<1$ are absolutely protected (do not appear in the decomposition of other multiplets).
The BPS condition for $B_1$ multiplets (\ref{eq:3dN4-multiplets}) matches the BPS condition for the geodesics in (\ref{eq:BPS-ab}), though $B_1$ and $A_2$ have $\Delta/J\rightarrow 1$ for large $J$.

\subsection{Janus ICFT}\label{sec:Janus-QFT}

The Janus ICFT realizes two 4d $\mathcal N=4$ SYM theories on half spaces, with independent gauge couplings $g^{}_{\rm YM,L}$ and $g^{}_{\rm YM,R}$, joined at an interface which does not host additional 3d degrees of freedom (fig.~\ref{fig:Janus-sugra}).
Half of the supersymmetry can be preserved by including an interface-localized interaction term in the Lagrangian, which takes the schematic form \cite[eq.~(2.2)]{DHoker:2007zhm}
\begin{align}\label{eq:Janus-interface-L}
	\mathcal L_{\rm interface}&=-\frac{2}{3}i (\partial_\perp g) g^3 \epsilon^{ijk} X^i[X^j,X^k]+{\rm fermions}~,
\end{align}
with $\partial_\perp$ the derivative perpendicular to the interface. This term singles out one triplet of scalar fields in the decomposition (\ref{eq:N4SYM-scalar-split}) and breaks the R-symmetry to $SO(3)\times SO(3)$.
For $g_{\rm YM,L}=g_{\rm YM,R}$ the interface term (\ref{eq:Janus-interface-L}) vanishes and the theory reduces to standard 4d $\mathcal N\,{=}\,4$ SYM.

From the holographic discussion in sec.~\ref{sec:AdS5xS5} we identify the following operators:
\begin{itemize}
	\item[--] One single-charge BPS operator with $\Delta=J_1$ and one with $\Delta=J_2$, which each give rise to a BMN-like sector with nearby operators described by 8 bosonic and 8 fermionic deformations, with scaling dimensions (\ref{eq:string-spectrum}) with
	\begin{align}\label{eq:Janus-spectrum}
		(\Delta-J_2)_n&=\sqrt{1+\frac{2g_{\rm YM,L}^2g_{\rm YM,R}^2}{g_{\rm YM,L}^2+g_{\rm YM,R}^2}\frac{n^2N}{J_2^2}}~,
		&
		(\Delta-J_1)_n&=\sqrt{1+\frac{g_{\rm YM,L}^2+g_{\rm YM,R}^2}{2}\frac{n^2N}{J_1^2}}~.
	\end{align}
	\item[--] One family of two-charge BPS operators with $|J_1|+|J_2|=\Delta$ and $(J_1,J_2)$ interpolating between $(\Delta,0)$ and $(0,\Delta)$, which each give rise to a sector described by a pp-wave limit.
\end{itemize}

For the identification of the BMN-like sectors we start with standard $\mathcal N=4$ SYM.
In \cite{Berenstein:2002jq} geodesics were placed at the origin of global $\rm AdS_5$, with the $S^5$ in $S^1\times S^3$ slicing and angular momentum on $S^1$.
The pp-wave vacuum emerging in the Penrose limit corresponds to a state in the CFT on the cylinder and by the state/operator map to an operator, identified in
\cite{Berenstein:2002jq} as $\tr Z^J$ with $Z=\phi^5+i\phi^6$.
This $\Delta=J$ BPS operator has charge $J$ under the $SO(2)$ rotating $(\phi^5,\phi^6)$ and is invariant under the $SO(4)$ subgroup of the $SO(6)$ $R$-symmetry acting on $(\phi^1,\phi^2,\phi^3,\phi^4)$.
String excitations were identified with the insertion of operators with $\Delta-J=1$ into $\tr Z^J$, with the momentum $n$ specifying phases included with the insertions.
Excitations of the 8 bosonic worldsheet fields correspond to insertions of $\phi^{1,2,3,4}$ and $D_\mu Z$, $\mu=0,1,2,3$,
and the fermionic modes correspond to inserting operators constructed from the 4 Weyl fermions of 4d $\mathcal N=4$ SYM.

In the $\rm AdS_5\times S^5$ coordinates (\ref{eq:AdS5S5-metric}) emerging from the $\rm AdS_4\times S^2\times S^2\times \Sigma$ realization, $\rm AdS_5$ in (global) $\rm AdS_4$ slicing realizes the cylinder on the boundary as union of two half cylinders.
The $S^2\times S^2$ slicing of $S^5$ corresponds to splitting the scalar fields as in (\ref{eq:N4SYM-scalar-split}).
We define operators
\begin{align}\label{eq:Janus-ops}
	\tr Z^J~,&\quad Z=Y^2+iY^3~,
	&
	\tr \tilde Z^J~,\quad\tilde Z=X^2+iX^3~,
\end{align}
$\tr Z^J$ is charged under $SO(3)_Y$ and invariant under $SO(3)_X$, $\tr\tilde Z^J$ is charged under $SO(3)_X$ and invariant under $SO(3)_Y$.
They are related by an $SO(6)$ transformation and each preserves an $SO(4)$, though splitting the scalars as in (\ref{eq:N4SYM-scalar-split}) only leaves an $SO(3)$ manifest. 
Following \cite{Berenstein:2002jq}, we identify (\ref{eq:Janus-ops}) as the single-charge BPS operators seeding the BMN sectors (\ref{eq:Janus-spectrum}) for the special case $\rm AdS_5\times S^5$.
Nearby operators obtained by $N_n$ insertions of $\Delta-J_i=1$ operators with appropriate phases, as outlined above, have scaling dimension (\ref{eq:string-spectrum}) with $(\Delta-J_i)_n$ in (\ref{eq:Janus-spectrum}) with $g_{\rm YM,L}=g_{\rm YM,R}$.

The Janus ICFT is connected to standard 4d $\mathcal N=4$ SYM by an exactly marginal deformation (in the 3d sense).
We expect the single-charge BPS seed operators to still take the form (\ref{eq:Janus-ops}), which were tailored to the $SO(3)_X\times SO(3)_Y$ decomposition of $SO(6)$. 
Comparing to (\ref{eq:3dN4-multiplets}), these operators can be identified as scalar primaries in the  3d $\mathcal N=4$ multiplets
\begin{align}
	\tr Z^J:&\quad B_1[0]_{\Delta=J}^{(2J,0)}~,
	&
	\tr \tilde Z^J:&\quad B_1[0]_{\Delta=J}^{(0,2J)}~.
\end{align}
These operators match the expectations for the pp-wave vacua emerging from the two single-charge geodesics in (\ref{eq:Janus-geod-boundary}), for which string theory on the associated pp-wave geometries leads to the spectrum of nearby operators (\ref{eq:Janus-spectrum}).
The Janus deformation with $g_{\rm YM,L}\neq g_{\rm YM,R}$ singles out one triplet of scalars in  (\ref{eq:Janus-interface-L}), lifting the degeneracy between the spectra of nearby operators in (\ref{eq:Janus-spectrum}).

We expect the identification of string fluctuations with insertions of $\Delta-J_i=1$ operators to also carry over from $\mathcal N=4$ SYM.
With the $SO(6)$ symmetry and some supersymmetry broken, one would expect some degeneracies in the spectrum of nearby operators to be lifted. However, the enhanced symmetry of the pp-wave limits shows that the full degeneracy persists, with 8 degenerate bosonic and 8 fermionic deformations for each of the two single-charge seed operators.

The family of two-charge BPS operators may be realized by operators of the schematic form $\tr (Z^{J_1} \tilde Z^{J_2})$.
In 4d $\mathcal N=4$ SYM these are $\frac{1}{4}$-BPS operators, studied e.g.\ in \cite{Ryzhov:2001bp,DHoker:2003csh,Beisert:2003tq},
in 3d $\mathcal N=4$ they correspond to $B_1[0]^{2J_1,2J_2}$ multiplets.
Operators of this form arise from $\tr Z^{J_1}$ upon repeatedly inserting $\tilde Z$, following \cite[sec.~4]{Berenstein:2002jq}. 
As long as the $\tilde Z$ are dilute, this is described by string excitations on the pp-wave vacuum associated with $\tr Z^{J_1}$.
But repeated insertions eventually lead beyond the dilute gas approximation of \cite{Berenstein:2002jq} and to ``backreacted" operators with $J_1$, $J_2$ of the same order.

\subsection{D3/D5 BCFT}\label{sec:D3D5-ops}

Semi-infinite D3-branes ending on D5-branes engineer 4d $\mathcal N=4$ SYM on a single half space with Nahm pole boundary conditions. 
With the 4d $\mathcal N=4$ vector multiplet decomposed as in (\ref{eq:4d-vector-split}),
ending D3-branes on D5-branes imposes vanishing Dirichlet boundary conditions
for the 3d vector multiplet and a Nahm pole boundary condition on the 3d hypermultiplet scalars $\vec{X}$,
\begin{align}\label{eq:D3D5-bc-0}
	F_{\mu\nu}\big\vert_{x_3=0}&=0~, &
	Y_i\big\vert_{x_3=0}&=0~,
	&
	D_3 X_i -\frac{i}{2}\epsilon_{ijk}[X_j,X_k]\big\vert_{x_3=0}&=0~.
\end{align}
The general solution for $\vec{X}$ involves a pole specified by a set of $SU(2)$ generators $t^i$,
\begin{align}\label{eq:D3D5-bc}
	X^i&\sim\frac{t^i}{x^3}~,
	&
	[t^i,t^j]&=i\epsilon^{ijk}t^k~.
\end{align}
How the D3-branes end on the D5-branes determines the representation of $t_i$.
For D3-branes ending on $N_5$ D5-branes with $K_j$ D3-branes ending on the $j^{\rm th}$ D5, the generators take the form
\begin{align}\label{eq:ti-decom-gen}
	t_i&=t_i^{K_1\times K_1}\oplus t_i^{K_2\times K_2}\oplus\ldots \oplus t_i^{K_{N_5}\times K_{N_5}}~,
\end{align}
where $t_i^{K_j\times K_j}$ denotes the $K_j$-dimensional irreducible representation. On the boundary there is no dynamical gauge field and a global symmetry emerges \cite[sec.~2.4]{Gaiotto:2008sa}.
The $t^i$ define an embedding of $\mathfrak{su}(2)$ into the Lie algebra of the gauge group, $\mathfrak{g}$, and the global symmetry arises from (constant) gauge transformations associated with the subalgebra $\mathfrak{f}$ of $\mathfrak{g}$ that commutes with this $\mathfrak{su}(2)$.

\begin{figure}
	\subfigure[][]{\label{fig:D3D5-brane-1}
		\begin{tikzpicture}[yscale=0.75]
			\pgfmathsetmacro{\s}{1}
			
			\foreach \i in {0,...,2} \draw[thick] (\i*\s,-1.5) -- +(0,3);
			\foreach \i in {-2.5,-1.5} \draw (0,0.2*\i) -- (4*\s,0.2*\i);
			\foreach \i in {-0.5,0.5} \draw (\s,0.2*\i) -- (4*\s,0.2*\i);
			\foreach \i in {1.5,2.5} \draw (2*\s,0.2*\i) -- (4*\s,0.2*\i);
		\end{tikzpicture}
	}
	\hskip 15mm
	\subfigure[][]{\label{fig:D3D5-brane-2}
		\begin{tikzpicture}[yscale=0.75]
			\pgfmathsetmacro{\s}{0.3}
			\pgfmathsetmacro{\t}{0.1}
			
			\foreach \i in {0,...,3} \draw[thick] (\i*\s,-1.5) -- +(0,3);
			\foreach \i in {5,...,7} \draw[thick] (\i*\s,-1.5) -- +(0,3);
			\foreach \i in {9,10} \draw[thick] (\i*\s,-1.5) -- +(0,3);
			
			\foreach \j in {0,1,2,3} \draw (\j*\s,-7.5*\t+\j*\t) -- (20*\s,-7.5*\t+\j*\t);
			
			\foreach \j in {5,6} \draw (5*\s,-7.5*\t+\j*\t) -- (20*\s,-7.5*\t+\j*\t);
			\foreach \j in {7,8} \draw (6*\s,-7.5*\t+\j*\t) -- (20*\s,-7.5*\t+\j*\t);
			\foreach \j in {9,10} \draw (7*\s,-7.5*\t+\j*\t) -- (20*\s,-7.5*\t+\j*\t);
			
			\foreach \j in {12,13,14} \draw (9*\s,-7.5*\t+\j*\t) -- (20*\s,-7.5*\t+\j*\t);
			\foreach \j in {15,16,17} \draw (10*\s,-7.5*\t+\j*\t) -- (20*\s,-7.5*\t+\j*\t);

		\end{tikzpicture}
	}
	\caption{Left: $N_5K$ D3-branes ending on $N_5$ 5-branes, with $K$ D3-branes ending on each 5-brane, for $N_5=3$, $K=2$. The vertical lines are D5-branes for D3/D5 and NS5-branes for D3/NS5. Right: $P$ groups of 5-branes, with $N_5^{(i)}$ 5-branes in each group and $K_i$ D3 ending on each 5-brane in that group, for $P=3$, $N_5^{(1)}=4$, $K_1=1$, $N_5^{(2)}=3$, $K_2=2$, $N_5^{(3)}=2$, $K_3=3$.}
\end{figure}
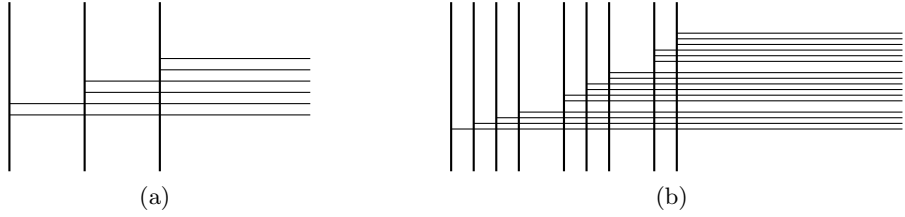

The D3/D5 BCFT as described by (\ref{eq:h1h2-D3D5-BCFT}) in sec.~\ref{eq:D3NS5-sugra} has $N_5 K$ D3-branes ending on $N_5$ D5-branes, with $K$ D3 ending on each D5 (fig.~\ref{fig:D3D5-brane-1}). 
The $t_i$ are block-diagonal $N_5K\,{\times}\, N_5K$ matrices with $K\,{\times}\, K$ blocks $t_i^{K\times K}$ on the diagonal,
\begin{align}\label{eq:ti-decom-N5K}
	t_i&=t_i^{K\times K}\oplus \ldots \oplus t_i^{K\times K}~.
\end{align}
The global symmetry arises in the brane setup from the $U(N_5)$ associated with the $N_5$ D5 branes. 
$K=1$ leads to a standard  Neumann boundary condition for the 3d hypermultiplet.
The holographic discussion in sec.~\ref{eq:D3NS5-sugra} identifies the following operators:
\begin{itemize}
	\setlength\itemsep{0.1em}
	\item[--] One single-charge BPS operator for $SO(3)_X$ which gives rise to a BMN-like sector of nearby operators described by 8 bosonic and 8 fermionic deformations with scaling dimensions
	\begin{align}\label{eq:D3D5-spectrum}
		(\Delta-J_1)_n&=\sqrt{1+\left(1+\frac{\cosh^{-1}( k+1)}{\sqrt{k(k+2)}}\right)\frac{n^2g_{4d}^2 N_5K}{4J_1^2}}~,
		&
		k&= \frac{4\pi^2}{g_{\rm 4d}^2}\frac{K}{N_5}~.
	\end{align}
	\item[--] For each $J_2/\Delta$ up to a maximal value ((\ref{eq:D3NS5-J1-BPS-max}), fig.~\ref{fig:D3NS5-J1max} with $J_1\rightarrow J_2$), 2 two-charge BPS operators with $|J_1|+|J_2|=\Delta$, which each give rise to a sector described by a pp-wave limit.
	
	\item[--] Evidence for two single-charge non-BPS operators, with $J_i/\Delta$ in (\ref{eq:D3NS5-single-charge-nonBPS}), (\ref{eq:D3NS5-single-charge-nonBPS-2}) with $J_1\leftrightarrow J_2$, and a family of two-charge non-BPS seed operators interpolating between them (fig.~\ref{fig:D3NS5-J-non-BPS}).
\end{itemize}

At the boundary of the D3/D5 BCFT, the ambient operators $\tr Z^J$ in (\ref{eq:Janus-ops}) are projected out by the boundary condition (\ref{eq:D3D5-bc-0}).
This is consistent with the absence of BPS geodesics with only $SO(3)_Y$ charge.
For $X^i$ the boundary condition permits dynamics on the background (\ref{eq:D3D5-bc}), described by string theory on the holographic dual. With $\hat X^i$ denoting the regular part of $X^i$ at $x^3=0$, so that $X^i=t^i/x^3+\hat X^i$, we can define a single-charge BPS boundary operator
\begin{align}\label{eq:D3D5-ambient}
	&\lim_{x^3\rightarrow 0}\tr \tilde Z^J~, & \tilde Z&=\hat X^2+i\hat X^3~.
\end{align}
As extrapolation of a gauge-invariant single-trace ambient operator it is invariant under the global symmetry associated with the boundary condition (\ref{eq:D3D5-bc-0}) and a candidate to seed the BMN-like sector (\ref{eq:D3D5-spectrum}).
Additional genuinely 3d operators with $\Delta=J_1$ can be formed from components of $\tilde Z$ which would not be gauge invariant in the ambient CFT but are permitted on the boundary \cite[sec.~2.2.2]{Gaiotto:2008sa}.
The singular nature of the Nahm pole imposes constraints on these operators: following \cite[sec.~5]{Dedushenko:2020vgd} the admissible boundary operators are the components of $\tilde Z$ in the subalgebra $\mathfrak{f}$ of the Lie algebra of the gauge group that emerges as global symmetry. 
These boundary operators generally transform under the non-Abelian global symmetry and we expect them to correspond to open strings at the D5-brane source. 
An invariant combination can be formed in the present example using the trace in $\mathfrak{f}=\mathfrak{u}(N_5)$, and this coincides (up to normalization) with (\ref{eq:D3D5-ambient}) as candidate seed operator.

The pairs of two-charge BPS seed operators with $J_2/\Delta$ up to (\ref{eq:D3NS5-J1-BPS-max})$\vert_{J_1\rightarrow J_2}$ are predictions.
With $\vec{Y}$ projected out by the boundary condition \eqref{eq:D3D5-bc-0} there are no obvious candidate BPS operators with only $SO(3)_Y$ charge in the free theory, from which we could define candidate two-charge BPS seed operators.
This matches the holographic results: for $g_{4d}\rightarrow 0$ at fixed $N_5$ and $K$, $k\rightarrow \infty$ and the maximal $SO(3)_Y$ charge vanishes (fig.~\ref{fig:D3NS5-J1max}$\vert_{J_1\rightarrow J_2}$). 
This suggests that the two-charge BPS seed operators for finite $k$ arise, upon increasing the coupling, in the decomposition of long multiplets hitting the BPS bound.
The non-BPS seed operators should be in long multiplets.

The discussion can be generalized to what we called D3/D5$^P$ in sec.~\ref{eq:D3NS5-sugra}, i.e.\ $P$ groups of D5-branes with the $i^{\rm th}$ group comprising $N_5^{(i)}$ D5-branes with $K_i$ D3-branes ending on each (fig.~\ref{fig:D3D5-brane-2}).
The $t^i$ in (\ref{eq:ti-decom-gen}) are block diagonal matrices with $N_5^{(r)}$ $K_r\times K_r$ blocks for $r=1,..,P$,
\begin{align}
	t_i&=\oplus_{r=1}^P\Big(\oplus_{s=1}^{N_5^{(r)}} t_i^{K_r\times K_r}\Big)~.
\end{align}
The global symmetry emerges from the $\prod_{r=1}^P U(N_5^{(r)})$ associated with the $P$ D5-brane groups, with $U(N_5^{(r)})$ transforming the $K_rN_5^{(r)}\times K_rN_5^{(r)}$ blocks.
The holographic dual has $P$ separated D5-brane sources on the boundary of $\Sigma$ (as for D3/NS5$^P$ in (\ref{eq:h1h2-D3NS5-BCFT-p})), and $P$ single-charge BPS geodesics, as shown in sec.~\ref{sec:null-geodesics}.\footnote{%
In the limit where all $K_r$ become equal the global symmetry in the BCFT enhances; in the supergravity dual the $P$ 5-brane sources merge and $P-1$ geodesics end up at this resulting 5-brane source.}
Each gives rise to a BMN-like sector for which the spectrum of nearby operators can be obtained by straightforwardly generalizing the discussion in sec.~\ref{eq:D3NS5-sugra}.
The boundary operators discussed above now comprise components of $\tilde Z$ in $\mathfrak{f}=\mathfrak{u}(N_5^{(1)})\oplus\ldots\oplus u(N_5^{(P)})$. For each $\mathfrak u(N_5^{(r)})$ an invariant combination can be formed as the trace, providing $P$ candidate seed operators, as desired. They may mix and one combination is equivalent to (\ref{eq:D3D5-ambient}).\footnote{In fig.~\ref{fig:D3D5-brane-2} the $P$ 5-brane groups can be separated without breaking the global symmetry, to resolve the boundary into a sequence of 4d $\mathcal N=4$ SYM on intervals. Analogs of (\ref{eq:Janus-ops}) can then be defined for each interval. Upon collapsing the intervals they give rise to $P-1$ operators in addition to (\ref{eq:D3D5-ambient}) for D3/D5$^P$ (and (\ref{eq:D3NS5-4d-op}) for D3/NS5$^P$).\label{foot:stretch}}

\subsection{D3/NS5 BCFT}\label{sec:D3NS5-ops}

Semi-infinite D3-branes ending on NS5 branes engineer 4d $\mathcal N\,{=}\,4$ SYM on a half space coupled to a 3d $\mathcal N\,{=}\,4$ SCFT on the boundary. This is the S-dual of the D3/D5 BCFTs.
The 4d vector multiplet splits as in (\ref{eq:4d-vector-split}).
Terminating D3-branes on a single NS5 imposes a Dirichlet boundary condition for the 3d hypermultiplet and a Neumann condition for the 3d vector multiplet \cite{Gaiotto:2008sa},
\begin{align}\label{eq:D3-single-NS5}
	\vec{X}\big\vert_{x^3=0}&=0~,
	&
	F_{3\mu}\big\vert_{x^3=0}=D_3\vec{Y}\big\vert_{x^3=0}=0~.
\end{align}
The 3d vector multiplet in (\ref{eq:4d-vector-split}) is dynamical and the full gauge symmetry is realized on the boundary.
For D3-branes ending on multiple NS5-branes the 3d vector multiplet in (\ref{eq:4d-vector-split}) couples to a 3d $T_\rho^\sigma[SU(N)]$ SCFT on the boundary.
These 3d SCFTs can be described as IR limit of 3d gauge theories where the (dimensionful) 3d gauge couplings become strong.
Terminating $N_5K$ D3-branes on $N_5$ NS5-branes such that $K$ D3-branes end on each NS5 (fig.~\ref{fig:D3NS5-brane-2}) engineers the quiver
\begin{align}\label{eq:D3NS5-quiver}
	U(K)-U(2K)-\ldots - U((N_5-1)K)-\widehat{U(N_5K)}
\end{align}
where the hatted node denotes 4d $\mathcal N=4$ SYM on a half space and all others are 3d nodes. The BCFT emerging as IR limit retains the 4d gauge coupling as marginal parameter.
It has a global symmetry realized in the brane construction (fig.~\ref{fig:D3D5NS5-brane}) by the $U(N_5)$ associated with the NS5 branes. 
The gauge theory realizes a topological $U(1)$ symmetry for each 3d vector multiplet with conserved current $\star\tr F_t$, which is enhanced at the IR fixed point by monopole operators \cite{Borokhov:2002ib,Borokhov:2002cg,Borokhov:2003yu,Cremonesi:2013lqa}.

From the holographic discussion in sec.~\ref{eq:D3NS5-sugra} we identify the following operators
\begin{itemize}
	\setlength\itemsep{0.1em}
	\item[--] One single-charge BPS operator with $\Delta=J_2$ which gives rise to a BMN-like sector of nearby operators described by 8 bosonic and 8 fermionic deformations with scaling dimensions
	\begin{align}\label{eq:D3NS5-spectrum}
		(\Delta-J_2)_n&=\sqrt{1+\left(\sqrt{k+2}+\frac{\cosh ^{-1}(k+1)}{\sqrt{k}}\right)^2\frac{n^2g_{4d}^2N_5K}{4J_2^2}}~,
		&
		k&=\frac{g_{4d}^2}{4}\frac{K}{N_5}~.
	\end{align}	
	\item[--] For each $J_1/\Delta$ up to a maximal value given in (\ref{eq:D3NS5-J1-BPS-max}) and fig.~\ref{fig:D3NS5-J1max}, 2 two-charge BPS operators with $|J_1|+|J_2|=\Delta$, which each give rise to a sector described by a pp-wave limit.
	
	\item[--] Evidence for two single-charge non-BPS operators with $J_i/\Delta$ in (\ref{eq:D3NS5-single-charge-nonBPS}), (\ref{eq:D3NS5-single-charge-nonBPS-2}) and a family of two-charge non-BPS seed operators interpolating between them (fig.~\ref{fig:D3NS5-J-non-BPS}).
	
\end{itemize}

The discussion of ambient operators mirrors the D3/D5 case. Of the operators in (\ref{eq:Janus-ops}) the NS5 boundary conditions retain the operator charged under $SO(3)_Y\sim SO(3)_C$,
\begin{align}\label{eq:D3NS5-4d-op}
	&\lim_{x^3\rightarrow 0}\tr^{}_{U(N_5K)} Z^J~,& Z&=Y^2+iY^3~,
\end{align}
which is invariant under the global symmetry.
This is one candidate to seed the BMN-like sector (\ref{eq:D3NS5-spectrum}), associated with the single-charge BPS geodesic in (\ref{eq:D3NS5-endpoints}).
Having $\tr\tilde Z^J$ in (\ref{eq:Janus-ops}) projected out is consistent with the absence of BPS geodesics with only $SO(3)_X$ charge.
Additional boundary-localized operators can be constructed from 3d fields, some of which we discuss below. The existence and properties of the two-charge BPS (and non-BPS) seed operators are predictions.

The discussion can again be generalized to D3-branes ending on $p$ groups of NS5-branes, with $N_5^{(i)}$ NS5-branes in the $i^{\rm th}$ group and $K_i$ D3-branes ending on each NS5-brane in the $i^{\rm th}$ group, as illustrated in fig.~\ref{fig:D3D5-brane-2} and described holographically by (\ref{eq:h1h2-D3NS5-BCFT-p}). 
The quiver becomes
{\small
	\begin{align}\label{eq:D5-NS5-P-quiver}
		\underbrace{U(L_{1}^{(1)})-\ldots-U(L^{(1)}_{N_5^{(1)}})}_\text{$N_5^{(1)}$ nodes}
		-
		\underbrace{U(L^{(2)}_{1})-\ldots-U(L^{(2)}_{N_5^{(2)}})}_\text{$N_5^{(2)}$ nodes}
		-
		\ldots
		-
		\underbrace{U(L^{(p)}_1)-\ldots-U(L^{(p)}_{N_5^{(p)}-1})}_\text{$N_5^{(p)}-1$ nodes}
		-
		\widehat{U(L^{(p)}_{N_5^{(p)}})}
\end{align}}%
where 
\begin{align}
	L_{a}^{(1)}&=aK_1~, & 
	L_a^{(i+1)}&=L_{N_5^{(i)}}^{(i)}+aK_{i+1}~.
\end{align}
The first $N_5^{(i)}-1$ nodes within each group are balanced, the nodes in-between overbalanced.
Each 3d vector multiplet contributes a topological $U(1)$ symmetry. This is enhanced in the IR limit by monopole operators so that each string of $N_5^{(i)}-1$ balanced 3d nodes gives rise to an $SU(N_5^{(i)})$ symmetry, while the in-between nodes contribute $U(1)$ symmetries \cite[sec.~2.4.3]{Gaiotto:2008ak}.
In the brane setup this corresponds to the $\prod_{i=1}^pU(N_5^{(i)})$ associated with the $p$ NS5-brane groups, with each factor acting on the NS5-branes corresponding to the string of $N_5^{(i)}-1$ balanced 3d nodes.

The holographic dual in (\ref{eq:h1h2-D3NS5-BCFT-p}) has $p$ NS5 sources and consequently $p$ single-charge BPS geodesics giving rise to BMN-like sectors, as shown in sec.~\ref{sec:geod-Penrose}. 
The spectrum of nearby operators can be determined as in sec.~\ref{eq:D3NS5-sugra}.
Additional boundary-localized candidate seed operators can be constructed from 3d fields.
The 3d vector multiplet for each gauge node in (\ref{eq:D3NS5-quiver}) contains a triplet of scalar fields. Denoting the triplet for the $t^{\rm th}$ node by  $\vec{\mathcal Y}_t$, the 3d nodes yield single-trace operators
\begin{align}\label{eq:NS5D3BCFT-3d-ops}
	&\tr \mathcal Z_t^J~, & \mathcal Z_t&=\mathcal Y_t^2+i\mathcal Y_t^3~,
\end{align}
with engineering dimension $J$.\footnote{Extrapolating them from the gauge theory to the IR BCFT amounts to an RG flow, and differs from the extrapolation from free to interacting theories along conformal lines for Janus and D3/D5.}
In the brane construction the 3d gauge nodes arise from 4d $\mathcal N=4$ SYM on intervals (the finite D3-brane segments in fig.~\ref{fig:D3NS5-brane-2}), and the operators (\ref{eq:NS5D3BCFT-3d-ops}) arise from analogs of (\ref{eq:D3NS5-4d-op}). So (\ref{eq:D3NS5-4d-op}) is a natural extension of this group.
From the discussion above, the $p-1$ 3d operators (\ref{eq:NS5D3BCFT-3d-ops}) associated with the unbalanced nodes, $t=N_5^{(i)}$ with $i=1,\ldots, p-1$ and the operator (\ref{eq:D3NS5-4d-op}) arising from 4d fields are invariant under the non-Abelian global symmetries.
They are candidate seed operators for the $p$ BMN-like sectors.\footnote{This aligns with the discussion of Wilson loops in \cite{Coccia:2021lpp,Uhlemann:2020bek}: The condition for single-charge BPS geodesics is identical to that for fundamental Wilson loops (sec.~\ref{sec:null-geodesics}). Fundamental Wilson loops associated with unbalanced 3d nodes are represented by strings away from 5-brane sources, those associated with balanced 3d nodes are at poles.}
In the brane setup this can be motivated by stretching out the D3-branes engineering the unbalanced nodes in (\ref{eq:D5-NS5-P-quiver}), to realize 4d $\mathcal N=4$ SYM on an interval without breaking up the $P$ NS5-brane groups, as in footnote \ref{foot:stretch}.

\subsection{D3/D5/NS5 BCFT}\label{sec:D3D5NS5-ops}

\begin{figure}
	\subfigure[][]{\label{fig:BCFT-brane-1}
		\begin{tikzpicture}[xscale=0.7,yscale=0.75]
			\pgfmathsetmacro{\s}{0.7}
			
			\draw[fill=black] (0,0)  ellipse (2pt and 5pt);	
			\draw[fill=black] (\s,0)  ellipse (2pt and 10.5pt);	
			\draw[fill=black] (2*\s,0)  ellipse (2pt and 16pt);	
			\draw[fill=black] (3*\s,0)  ellipse (2pt and 21pt);	
			\draw[fill=black] (4*\s,0)  ellipse (2pt and 26pt);	
			\draw[fill=black] (6*\s,0)  ellipse (2pt and 26pt);
			\draw[fill=black] (7*\s,0)  ellipse (2pt and 23pt);
			\draw[fill=black] (8*\s,0)  ellipse (2pt and 20pt);
			
			\foreach \i in {-3.5,...,3.5} \draw (5*\s+0.1*\i,-1.5) -- +(0,3);
			
			\foreach \i in {-9,...,9} \draw (7*\s,0.075*\i) -- +(\s,0);
			\foreach \i in {-10.5,...,10.5} \draw (6*\s,0.075*\i) -- +(\s,0);
			\foreach \i in {-12,...,12} \draw (4*\s,0.075*\i) -- +(2*\s,0);
			\foreach \i in {-9.5,...,9.5} \draw (3*\s,0.075*\i) -- +(\s,0);
			\foreach \i in {-7,...,7} \draw (2*\s,0.075*\i) -- +(\s,0);
			\foreach \i in {-4.5,...,4.5} \draw (\s,0.075*\i) -- +(\s,0);
			\foreach \i in {-2,...,2} \draw (0,0.075*\i) -- +(\s,0);
			
			\foreach \i in {-7.5,...,7.5} \draw (8*\s,0.075*\i) -- +(8.5-8*\s,0);
		\end{tikzpicture}
	}
	\hskip 3mm
	\subfigure[][]{\label{fig:BCFT-brane-3}
		\begin{tikzpicture}[xscale=0.7,yscale=0.75]
			\pgfmathsetmacro{\s}{0.7}
			
			\draw[fill=black] (0,0)  ellipse (2pt and 4pt);	
			\draw[fill=black] (\s,0)  ellipse (2pt and 9pt);	
			\draw[fill=black] (2*\s,0)  ellipse (2pt and 14pt);	
			\draw[fill=black] (3*\s,0)  ellipse (2pt and 18pt);	
			
			\foreach \i in {-5.5,...,5.5} \draw (2*\s,0.075*\i) -- +(\s,0);
			\foreach \i in {-3.5,...,3.5} \draw (\s,0.075*\i) -- +(\s,0);
			\foreach \i in {-1.5,...,1.5} \draw (0,0.075*\i) -- +(\s,0);
			
			\foreach \i in {-7.5,...,7.5} \draw (3*\s,0.075*\i) -- (6,0.075*\i);
			
			\foreach \i in {-1.5,...,1.5}{
				\draw (4*\s+0.1*\i-0.2,-1.5) -- +(0,3);
			}
			
		\end{tikzpicture}
	}
	\hskip 3mm
	\subfigure[][]{\label{fig:BCFT-brane-2}
		\begin{tikzpicture}[xscale=0.7,yscale=0.75]
			\pgfmathsetmacro{\s}{0.7}
			
			\draw[fill=black] (0,0)  ellipse (2pt and 5pt);	
			\draw[fill=black] (\s,0)  ellipse (2pt and 10.5pt);	
			\draw[fill=black] (2*\s,0)  ellipse (2pt and 16pt);	
			\draw[fill=black] (3*\s,0)  ellipse (2pt and 21pt);	
			
			\foreach \i in {-7,...,7} \draw (2*\s,0.075*\i) -- +(\s,0);
			\foreach \i in {-4.5,...,4.5} \draw (\s,0.075*\i) -- +(\s,0);
			\foreach \i in {-2,...,2} \draw (0,0.075*\i) -- +(\s,0);
			
			\foreach \i in {-9.5,...,9.5} \draw (3*\s,0.075*\i) -- (6,0.075*\i);
			
			\foreach \i in {-1.5,...,1.5}{
				\draw (4*\s-0.1*\i,{0.075*(\i-12)}) -- (6,{0.075*(\i-12)});
				\draw (4*\s+0.1*\i,-1.5) -- +(0,3);
			}
			
		\end{tikzpicture}
	}
	\caption{The brane setup in fig.~\ref{fig:D3D5NS5-brane} after Hanany-Witten transitions: for $N_5>2K$ (left), $N_5=2K$ (center) and $N_5<2K$ (right), with NS5-branes as ellipses, D3/D5-branes as horizontal/vertical lines.\label{fig:D3D5NS5-brane-2}}
\end{figure}
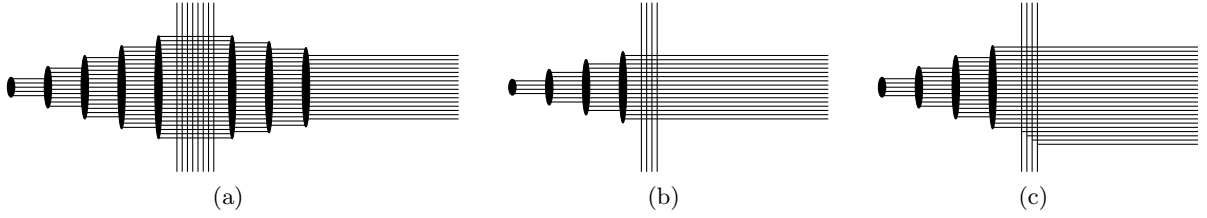

The D3/D5/NS5 BCFT is realized by $2N_5K$ D3-branes ending on $N_5$ D5-branes and $N_5$ NS5-branes (fig.~\ref{fig:D3D5NS5-brane}). The gauge theory description depends on $K/N_5$ and was discussed in \cite[sec.~4.5]{Coccia:2021lpp}, \cite{Karch:2022rvr}. 
For $N_5>2K$ (fig.~\ref{fig:BCFT-brane-1}) the quiver is
\begin{align}\label{eq:D5NS5K-quiver-2}
	U(R)-U(2R)-\ldots - &U(R^2) - U(R^2-S)-\ldots - U(2N_5K+S) - \widehat{U(2N_5K)}
	\nonumber\\
	&\ \ \ \vert\\
	\nonumber & \ [N_5]	
\end{align}
where $R=\frac{N_5}{2}+K$, $S=\frac{N_5}{2}-K$. The D5-branes add fundamental hypermultiplets to the 3d quiver.
For $N_5<2K$, shown in fig.~\ref{fig:BCFT-brane-2}, the D5-branes impose Nahm pole boundary conditions on part of the 4d fields, with $N_5|S|$ D3-branes ending on the $N_5$ D5-branes. This leaves a $U(N_5R)$ of the 4d $U(2N_5K)$ gauge group which couples to a 3d SCFT, as in the general construction in \cite[sec.~2.5]{Gaiotto:2008sa}. The combined 3d/4d quiver is denoted as
\begin{align}\label{eq:D5NS5K-quiver-1-rep}
	U(R)-U(2R)-\ldots - U((N_5-1)R) - \widehat{U(2N_5K)}
\end{align}
All 3d nodes are balanced in both quivers, and S-duality maps the theories to themselves.

In the holographic discussion we focused for simplicity on the case of self-dual 4d coupling, $g_{4d}^2=4\pi$. 
The 4d coupling can be restored using $SL(2,\ZZ)$, as spelled out in  \cite{Karch:2022rvr}.
We find
\begin{itemize}
	\setlength\itemsep{0.1em}
	\item[--] For $K\neq 0$ one single-charge BPS operator with $\Delta=J_1$ and one single-charge BPS operator with $\Delta=J_2$, which each give rise to a BMN-like sector of nearby operators described by 8 bosonic and 8 fermionic deformations with scaling dimensions
	\begin{subequations}\label{eq:D3D5NS5-spectrum}
	\begin{align}\label{eq:D3D5NS5-spectrum-1}
		(\Delta-J_2)_n&=\sqrt{1+\left(\sqrt{k (k+2)}+\cosh^{-1}(k+1)\right)^2\frac{n^2N_5^2}{J_2^2}}
		~,
		&
		k&=\frac{\pi K}{N_5}~,
		\\
		\label{eq:D3D5NS5-spectrum-2}
		(\Delta-J_1)_n&=\sqrt{1+k\frac{(k+2)^2}{(k+1)^2}\left(1+\frac{\cosh ^{-1}(k+1)}{\sqrt{k (k+2)}}\right)\frac{n^2N_5^2}{J_1^2}}
	\end{align}
	\end{subequations}
	
	\item[--] The spectrum of two-charge BPS seed operators depends on $k$ (fig.~\ref{fig:D3D5NS5-J1J2-3}): For $k<k_\star$, with $k_\star=\sqrt{2}-1$,
	 there is a pair of operators for each $0<J_1/\Delta=1-J_2/\Delta<1$.
	For $k>k_\star$ an intermediate `gap' region of $J_1/\Delta$ values is excluded.
	
	\item[-] Two single-charge BPS seed operators, one for $SO(3)_X$ and one for $SO(3)_Y$, with $J_i/\Delta$ in fig.~\ref{fig:D3D5NS5-J1J2-nonBPS-single}, and a family of two-charge non-BPS seed operators interpolating between them.
\end{itemize}

For the discussion of seed operators we start with $N_5<2K$ shown in fig.~\ref{fig:BCFT-brane-2} and the quiver (\ref{eq:D5NS5K-quiver-1-rep}).
The ambient operators (\ref{eq:Janus-ops}) take the form
\begin{align}\label{eq:D3D5NS5-ops}
	\tr^{}_{U(2N_5K)} Z^J~,&\quad Z=Y^2+iY^3~,
	&
	\tr^{}_{U(2N_5K)} \tilde Z^J~,\quad\tilde Z=X^2+iX^3~.
\end{align}
Both give rise to non-trivial boundary operators: $\vec{X}$ and $\vec{Y}$ split into blocks with NS5-like boundary conditions and D5 boundary conditions, in line with fig.~\ref{fig:BCFT-brane-2}, so neither is projected out entirely. 
Both are invariant under the $U(N_5)$ global symmetries arising from the D5 and NS5 branes, and following the D3/D5 and D3/NS5 discussions they are candidates for the BMN sectors (\ref{eq:D3D5NS5-spectrum}).

For $N_5=2K$, shown in fig.~\ref{fig:BCFT-brane-3}, the D5-branes have no D3-branes ending on them.
They add 3d hypermultiplets in the fundamental representation of the 4d gauge group. The quiver is similar to (\ref{eq:D5NS5K-quiver-2}) but with the 3d fundamental hypermultiplets at the 4d node.
$\vec{X}$ and $\vec{Y}$ do not decompose into blocks with D5 and NS5 boundary conditions and in the decomposition of the 4d fields (\ref{eq:4d-vector-split}) the full 3d vector multiplet is dynamical. Similar to D3/NS5, the D3/D5/NS5 theory retains $\lim_{x^3\rightarrow 0}\tr Z^J$ as boundary operator.
The 3d hypermultiplets at the 4d node modify the NS5 boundary condition for $\vec{X}$, following \cite[sec.~2.5]{Gaiotto:2008sa}, from vanishing Dirichlet as in (\ref{eq:D3-single-NS5}) to
\begin{align}
	\vec{X}\vert_{x^3=0}+\vec{\mu}&=0~,
\end{align}
where $\vec{\mu}$ is the moment map operator for the 3d hypermultiplets (the scalar primary in the current multiplet of the associated flavor symmetry).
The theory again retains $\lim_{x^3\rightarrow 0}\tr \tilde Z^J$ as non-trivial boundary operator as well, providing two candidate seed operators for the BMN-like sectors (\ref{eq:D3D5NS5-spectrum}).

Finally, for $N_5>2K$ (fig.~\ref{fig:BCFT-brane-1}) the quiver takes the form (\ref{eq:D5NS5K-quiver-2}). All 4d fields  get (generalized) NS5 boundary conditions and in the decomposition of the 4d fields (\ref{eq:4d-vector-split}) the full 3d vector multiplet is dynamical. This retains $\lim_{x^3\rightarrow 0}\tr Z^J$ in (\ref{eq:D3D5NS5-ops}) as boundary operator and candidate to seed the BMN-like sector (\ref{eq:D3D5NS5-spectrum-1}).
The 3d hypermultiplet in (\ref{eq:4d-vector-split}), on the other hand, is projected out at $x^3=0$.
To identify a candidate operator for (\ref{eq:D3D5NS5-spectrum-2}), we discuss the symmetries in more detail. The flavor symmetry splits into $G_C\times G_H$, realized by current multiplets with bottom components charged under $SO(3)_C$ and $SO(3)_H$, respectively.
Holographically, the gauge bosons dual to $G_H$ currents arise from $F$-strings and those for $G_C$ currents from D-strings, e.g.\ \cite[Table 1]{Bachas:2017wva}.

The BCFT (\ref{eq:D5NS5K-quiver-2}) couples 4d $\mathcal N=4$ SYM to a 3d SCFT. The 3d SCFT can be isolated by terminating each semi-infinite D3-brane in fig.~\ref{fig:BCFT-brane-1} on a separate D5, as in \cite{Karch:2022rvr}. This replaces the 4d gauge node in (\ref{eq:D5NS5K-quiver-2}) with a 3d flavor node. The 3d SCFT has $G_C=SU(N_5)$, emerging from topological $U(1)$ symmetries enhanced by monopole operators, and $G_H=S(U(N_5)\times U(2N_5K))$, with the $N_5$ and $2N_5K$ flavors contributing $U(N_5)$ and $U(2N_5K)$, respectively, and an overall $U(1)$ decoupling \cite[sec.~2.4.3]{Gaiotto:2008ak}. 
When coupling this 3d SCFT to 4d $U(2N_5K)$ $\mathcal N=4$ SYM on a half space, the 4d fields gauge the 3d $SU(2N_5K)$ flavor symmetry, while the $U(1)$ in $U(2N_5K)$ terminates with a Neumann boundary condition.
The 3d vector multiplet arising from the 4d fields contributes a $U(1)$ current for $G_C$. The flavor symmetry then has the (local) form $G_C=G_H=SU(N_5)\times U(1)$.

Following this discussion we focus on $G_H$ and single out the $U(1)$, for which the associated moment map operator is invariant under the global symmetry. Denoting it by $\vec{\mu}$, we define
\begin{align}\label{eq:D3D5NS5-moment-op}
	&\tr^{}_{U(R^2)} \hat Z^J~,& \hat Z&=\mu^2+i\mu^3~.
\end{align}
This is a candidate seed operator with only $SO(3)_X$ charge for the BMN-like sector in (\ref{eq:D3D5NS5-spectrum-2}).

This identification of candidate seed operators is consistent with the limit $K\rightarrow 0$ (which implies $N_5>2K$).
For $K= 0$ the 4d ambient CFT becomes trivial and the BCFT (\ref{eq:D5NS5K-quiver-2}) reduces to a genuine 3d CFT. The ambient operator in (\ref{eq:D3D5NS5-ops}) disappears. The flavor symmetry reduces to $G_C=G_H=SU(N_5)$ and the operator in (\ref{eq:D3D5NS5-moment-op}) likewise disappears. 
Correspondingly, the geodesics giving rise to both BMN-like sectors (\ref{eq:D3D5NS5-spectrum}) disappear in the holographic dual.

\subsection{\texorpdfstring{D5$^2$/NS5$^2$}{D52/NS52} interface}\label{sec:D52NS52-ops}

The ICFT dubbed D5$^2$/NS5$^2$ interface in \cite{Uhlemann:2023oea} is realized by D3-branes intersecting 2 groups of D5-branes and 2 groups of NS5-branes (fig.~\ref{fig:D52NS52-brane} with D3-brane numbers in (\ref{eq:D52NS52-ND3})). The gauge theory description depends on $\Delta_{k,\delta}$, with similar case distinctions as for D3/D5/NS5. We focus on $\Delta_{k,\delta}>0$ where the quiver, with $s=N_5\Delta_{k,\delta} /2$, is 
\begin{align}\label{eq:D52-NS52-quiver}
	\widehat{(N_{\rm D3}^\infty)}- (N_{\rm D3}^\infty+s)-\ldots - (&N_{\rm D3}^\infty+s^2) - \ldots - (X) - \ldots - (N_{\rm D3}^\infty+s^2) - \ldots - (N_{\rm D3}^\infty+s) - \widehat{(N_{\rm D3}^\infty)}
	\nonumber \\
	&\quad\ \vert\  \qquad  \qquad \qquad \qquad \quad \quad   \qquad\quad \vert
	\nonumber\\
	& \left[N_5/2\right]	\qquad \qquad \quad\qquad \qquad \qquad
	\left[N_5/2\right]
\end{align}%
Along the first ellipsis the ranks increase in steps of $s$, along the second ellipsis they decrease in steps of $N_5/2-s$. The central node with $X=N_{\rm D3}^\infty+\frac{1}{4}N_5^2(2\Delta_{k,\delta}-1)$ is overbalanced, and the quiver is symmetric under reflection across this node.
For $K=0$ the ambient CFTs become trivial, and the ICFT reduces to a 3d CFT. This 3d CFT was used for a string theory realization of wedge holography in \cite{Uhlemann:2021nhu} and results for free energy and Wilson loops were obtained in \cite{Coccia:2020wtk,Coccia:2021lpp}. 

In the holographic discussion in sec.~\ref{sec:D52-NS52-sugra} we set the 4d gauge couplings to self-dual values on both sides of the interface. The field theory then has a reflection symmetry $x^3\rightarrow -x^3$. We focused on BPS operators and found
\begin{itemize}
	\setlength\itemsep{0.1em}
	\item[--] 3 single-charge BPS seed operators with $\Delta=J_2$, each seeding a BMN-like sector with $\Delta-J_2$ as in (\ref{eq:string-spectrum}): one invariant under $x^3\rightarrow -x^3$ with $\tilde f_4^2$ in (\ref{eq:D52-NS52-interface-f4t-1a}), and for $K>0$ an additional pair of seed operators exchanged under $x^3\rightarrow -x^3$ with $\tilde f_4^4$ in (\ref{eq:D52-NS52-interface-f4t-2a})
	
	\item[--] 3 single-charge BPS seed operators with $\Delta=J_1$, each seeding a BMN-like sector with $\Delta-J_1$ as in (\ref{eq:string-spectrum}): one invariant under $x^3\rightarrow -x^3$ with $\tilde f_4^2$ in (\ref{eq:D52-NS52-interface-f4t-1b}), and for $K>0$ an additional pair of seed operators exchanged under $x^3\rightarrow -x^3$ with $\tilde f_4^4$ in (\ref{eq:D52-NS52-interface-f4t-2b})
	
	\item[--] A one-parameter family of 2-charge BPS seed operators interpolating between the single-charge operators (\ref{eq:D52-NS52-interface-f4t-1a}) and (\ref{eq:D52-NS52-interface-f4t-1b})
	
	\item[--] For $K>0$ four additional families of 2-charge BPS seed operators for whom the range of $J_i/\Delta$ depends on $k$ and $\delta$, with critical values in (\ref{eq:D52-NS52-k-delta-crit}).
	For $K=0$ two additional families, with $\delta_\star$ in (\ref{eq:delta-crit}) separating a regime where any $J_1/J_2$ is realized from a regime with a gap.
\end{itemize}

The discussion of field theory operators is similar to the D3/D5/NS5 quiver in (\ref{eq:D5NS5K-quiver-2}), fig.~\ref{fig:BCFT-brane-1}.
The 4d ambient fields on both sides in (\ref{eq:D52-NS52-quiver}) get generalized NS5 boundary conditions, and in the decomposition of the 4d fields (\ref{eq:4d-vector-split}) the full 3d vector multiplets are dynamical.
For this ICFT the limits from the left and right are independent, and two single-charge operators emerge from (\ref{eq:Janus-ops}),
\begin{align}\label{eq:D52-NS52-ambient-ops-1}
		&\lim_{x^3\rightarrow 0^+}\tr Z^J~, &
		&\lim_{x^3\rightarrow 0^-}\tr Z^J~,& 
		Z&=Y^2+iY^3~.
\end{align}
The reflection $x^3\,{\rightarrow}\, -x^3$ exchanges them.
Additional 3d operators with only $SO(3)_Y$ charge include
\begin{align}\label{eq:D52-NS52-3d-ops}
	&\tr \mathcal Z_t^J~, & \mathcal Z_t&=\mathcal Y_t^2+i\mathcal Y_t^3~,
\end{align}
where $t=1,..,2N_5-1$ labels the 3d gauge nodes in (\ref{eq:D52-NS52-quiver}).
Similar to the D3/NS5$^P$ and D3/D5/NS5 discussions, 
each string of $N_5/2-1$ balanced 3d gauge nodes in (\ref{eq:D52-NS52-quiver}) gives rise to an $SU(N_5/2)$ global symmetry, while the central node contributes a $U(1)$. The operator (\ref{eq:D52-NS52-3d-ops}) associated with the central 3d node $t=N_5$ is invariant under the non-Abelian global symmetry.
It is also invariant under reflection of $x^3$.

The pair of operators in (\ref{eq:D52-NS52-ambient-ops-1}) and (\ref{eq:D52-NS52-3d-ops})$\vert_{t=N_5}$ are candidate seed operators for the BMN-like sectors (\ref{eq:D52-NS52-interface-f4t-2a}), which are exchanged by $x^3\rightarrow -x^3$, and (\ref{eq:D52-NS52-interface-f4t-1a}), which is invariant. 
As we may define multiple combinations with the same transformation under the global symmetries, they may mix.

For the identification of candidate $SO(3)_X$ single-charge operators we again discuss the $G_H$ symmetries. Decoupling the 4d nodes in (\ref{eq:D52-NS52-quiver}) by terminating each semi-infinite D3-brane on a separate D5 leads to a 3d SCFT with $G_H=S(U(N_{\rm D3}^\infty)^2\times U(N_5/2)^2)$, emerging from the two groups of $N_5/2$ flavor hypermultiplets and two groups of $N_{\rm D3}^\infty$ flavors. 
Coupling the 3d SCFT to two 4d $\mathcal N=4$ SYM theories on half spaces gauges two $SU(N_{\rm D3}^\infty)$, and $G_H$ becomes (locally) $SU(N_5/2)^2\times U(1)^3$. The three operators defined from the moment maps for the $U(1)$ factors, $\vec{\mu}_i$, 
\begin{align}\label{eq:D52NS52-moment-op}
	&\tr^{} \hat Z_i^J~,& \hat Z_i&=\mu_i^2+i\mu_i^3~, \quad i=1,2,3,
\end{align}
are invariant under the global symmetry. They are candidate seed operators for the BMN-like sectors (\ref{eq:D52-NS52-interface-f4t-1b}) and (\ref{eq:D52-NS52-interface-f4t-2b}).

The candidates defined above are consistent with the limit $K\rightarrow 0$, which reduces the D5$^2$/NS5$^2$ ICFT to a 3d CFT. In this limit the $SO(3)_Y$ operators (\ref{eq:D52-NS52-ambient-ops-1}) disappear, while (\ref{eq:D52-NS52-3d-ops})$\vert_{t=N_5}$ remains. The $G_H$ flavor symmetry reduces to $S(U(N_5/2)^2)$, reducing the number of $U(1)$ moment-map operators in (\ref{eq:D52NS52-moment-op}) to one. 
In the holographic description, the geodesics in (\ref{eq:D52-NS52-interface-f4t-2a}), (\ref{eq:D52-NS52-interface-f4t-2b}) disappear, likewise leaving one BMN-like sector for $SO(3)_X$ and one for $SO(3)_Y$.

The candidates are also consistent with $\delta\rightarrow 0$. In this limit the groups of D5 and NS5 branes in fig.~\ref{fig:D52NS52-brane} merge to form one group of each. The quiver (\ref{eq:D52-NS52-quiver}) simplifies and the central node becomes balanced. 
The global symmetry enhances and the operator (\ref{eq:D52-NS52-3d-ops})$\vert_{t=N_5}$ is not invariant under the enhanced symmetry anymore, leaving the two operators in (\ref{eq:D52-NS52-ambient-ops-1}).
The number of invariant moment-map operators in (\ref{eq:D52NS52-moment-op}) likewise reduces to two.
In the holographic description the geodesics associated with (\ref{eq:D52-NS52-interface-f4t-1a}), (\ref{eq:D52-NS52-interface-f4t-1b}) end up at the poles, likewise leaving two single-charge geodesics at regular points of $\Sigma$ for $SO(3)_X$ and two for $SO(3)_Y$.

\section{Discussion}\label{sec:discussion}

We studied 4d $\mathcal N=4$ SYM with $\tfrac{1}{2}$-BPS boundaries, defects and interfaces, without assuming any quenched or probe limit. 
For each theory we identified sectors of operators described by string theory on pp-waves, with seed operators with various $SO(3)\times SO(3)$ charges
in short and long multiplets. 
The numbers and features of the seed operators depend on the theory.

Remarkably, for single-charge BPS seed operators the associated BMN-like sectors have the same maximal (super)symmetry as the BMN limit of standard 4d $\mathcal N=4$ SYM.
We showed that the number of such sectors is determined by the global symmetry of the CFT, 
and that this number can be arbitrarily large. Moreover, the spectrum of nearby operators in general differs for different BMN-like sectors within a single theory.
For a sample of B/I/dCFTs we explicitly derived the spectrum of nearby operators and discussed the field theory realization of the seed operators.

For the simplest case, the Janus interface, we identified two single-charge BPS seed operators which are analogous to the seed operator for 4d $\mathcal N=4$ SYM identified in \cite{Berenstein:2002jq}. The spectra of nearby operators (\ref{eq:Janus-spectrum}) are modified by the Janus deformation. We also included theories which are more drastic modifications of 4d $\mathcal N=4$ SYM, with boundaries and interfaces hosting genuine 3d degrees of freedom, engineered by D3-branes intersecting and/or ending on D5 and NS5 branes. 
In the brane constructions the 5-branes can be organized in groups such that each 5-brane within a group has the same net number of D3-branes ending on it. In the CFT each such 5-brane group gives rise to a global symmetry, and the number of groups determines the number of BMN-like sectors arising from single-charge BPS geodesics, $C_{1/2}$ in (\ref{eq:pp-wave-ops-gen}). 
This counting agrees with the number of $U(1)$ factors in the global symmetry.
The concrete spectra are summarized in sec.~\ref{sec:fieldtheory}, e.g.\ (\ref{eq:D3D5-spectrum}) for D3/D5 BCFTs, (\ref{eq:D3NS5-spectrum}) for D3/NS5, (\ref{eq:D3D5NS5-spectrum}) for the D3/D5/NS5 BCFT of \cite{Uhlemann:2021nhu} and for the D5$^2$/NS5$^2$ ICFT of \cite{Uhlemann:2023oea} in sec.~\ref{sec:D52NS52-ops},
along with candidate seed operators which in general involve, in addition to defect limits of ambient operators, also genuinely 3-dimensional operators.

We further identified seed operators for pp-wave sectors in unprotected short multiplets, whose existence is non-trivial information. 
They comprise families of operators which interpolate between single-charge BPS seed operators and cover certain ranges of $(J_1/\Delta,J_2/\Delta)$ which depend on the ambient 4d gauge couplings and parameters in the 3d CFTs.
We also found indications for seed operators in long multiplets for specific ratios of the charges and scaling dimensions.

Various questions are left for the future: 
We identified BMN-like sectors holographically and discussed the field theory realization of candidate seed operators. It would be desirable to fully identify the field theory realization of the BMN-like sectors, including the nearby operators described by string fluctuations.
Further tasks include deriving the spectrum of nearby operators for the two-charge BPS and non-BPS pp-wave limits, including an investigation of stability for the latter, and understanding the field theory realization.
Another direction is to generalize the geodesics to operators with spacetime spin, as in \cite{Gubser:2002tv}, and/or allow the geodesics to move on $\Sigma$. Returning to the most symmetric of operators, one may wonder whether fully backreacted solutions for the states may be constructed as extension of \cite{Lin:2004nb} for standard 4d $\mathcal N=4$ SYM.

It would also be desirable to connect the results derived here to integrability \cite{deLeeuw:2015hxa,Buhl-Mortensen:2015gfd,deLeeuw:2017cop,Buhl-Mortensen:2017ind,Dedushenko:2020yzd,deLeeuw:2024qki}, localization \cite{Robinson:2017sup,Dedushenko:2018tgx,Wang:2020seq,Komatsu:2020sup,Bason:2023bin} and bootstrap \cite{Liendo:2012hy,Liendo:2016ymz,deLeeuw:2017dkd,Baerman:2024tql,Chang:2019dzt} investigations of $\mathcal N=4$ SYM with boundaries and interfaces (and perhaps 3d $\mathcal N=4$ bootstrap more generally, e.g.~\cite{Chang:2019dzt}).
We note that the theories considered here are described by fully backreacted solutions; they are not described by probe D5-branes in $\rm AdS_5\times S^5$.
We also note that the 3d SCFTs appearing on boundaries and defects here are examples of long-quiver CFTs, i.e.\ theories whose planar limit involves a large number of gauge nodes.
This limit has been investigated in 5d \cite{Fluder:2018chf,Uhlemann:2019ypp} and 3d \cite{Coccia:2020cku,Coccia:2020wtk} field theories, with more recent works including \cite{Akhond:2021ffz,Akhond:2022oaf,Nunez:2023loo,Beccaria:2023qnu,Sobko:2024ohd}, and Penrose limits for such theories have been studied in \cite{Gutperle:2021nkl,Passias:2023meo}.

\let\oldaddcontentsline\addcontentsline
\renewcommand{\addcontentsline}[3]{}
\begin{acknowledgments}
	Part of this work was completed at the Aspen Center for Physics, which is supported by National Science Foundation grant PHY-2210452.
\end{acknowledgments}
\let\addcontentsline\oldaddcontentsline

\appendix

\section{Penrose limit details}\label{app:geod-details}

In this appendix we derive the results quoted in sec.~\ref{sec:null-geodesics} for geodesics and in \ref{sec:Penrose} for Penrose limits.
To simplify notation the we will omit the tilde for the 10d string frame metric functions.

\textbf{Geodesics:}
We seek null geodesics in the geometry \eqref{eq:met1} with (\ref{eq:ads4-global}), i.e.\ curves $x(\lambda)$ satisfying
\begin{align} 
	\frac{d^2x^\mu}{d\lambda^2}+\Gamma^{\mu}_{\nu\rho}\frac{dx^\nu}{d\lambda}\frac{dx^\rho}{d\lambda}&=0~,
	&
	g_{\mu\nu}\frac{dx^\mu}{d\lambda}\frac{x^\nu}{d\lambda}&=0~,
	\label{eq:geod-null}
\end{align}
with affine parameter $\lambda$.
We choose geodesics located at the equator of $S_1^2$ and $S_2^2$ and the center of $AdS_4$. This corresponds to $\theta_1=\theta_2=\hat{\rho}=0$.
The five resulting geodesic equations are:
\begin{align}
	\frac{d}{d\lambda}\left(\frac{dt}{d\lambda}f_4^2(z,\bar{z})\right)=0 
	\qquad\qquad
	\frac{d}{d\lambda}\left(\frac{d\phi_i}{d\lambda}f_i^2(z,\bar{z})\right)&=0
	\label{eq:geo1}
	\\[3mm]
	\frac{d^2z}{d\lambda^2} +\frac{\partial_{\bar{z}}f_4^2}{4\rho^2}\left(\frac{dt}{d\lambda}\right)^2-\frac{\partial_{\bar{z}}f_1^2}{4\rho^2} \left(\frac{d\phi_1}{d\lambda}\right)^2 -\frac{\partial_{\bar{z}}f_2^2}{4\rho^2} \left(\frac{d\phi_2}{d\lambda}\right)^2+\frac{\partial_z\rho^2}{\rho^2}\left(\frac{dz}{d\lambda}\right)^2&=0 \label{eq:geo4}
\end{align}
where $i=1,2$ and we note that the last equation is complex.
Equations  \eqref{eq:geo1} lead to the conserved quantities
\begin{align}
	\mathcal E &=f_4^2t'~, & \mathcal J_i&=f_i^2\phi_i^\prime~.
\end{align}
The null condition in (\ref{eq:geod-null}) and the remaining geodesic equation then become
\begin{align}
	-\frac{\mathcal E^2}{f_4^2}+\frac{\mathcal J_1^2}{f_1^2}+\frac{\mathcal J_2^2}{f_2^2}+4\rho^2|z'|^2&=0~,
	\label{eq:null2}
	\\
	z''+\frac{\partial_{\bar{z}}f_4^2}{4\rho^2}\left(\frac{\mathcal E}{f_4^2}\right)^2-\frac{\partial_{\bar{z}}f_1^2}{4\rho^2} \left(\frac{\mathcal J_1}{f_1^2}\right)^2 -\frac{\partial_{\bar{z}}f_2^2}{4\rho^2} \left(\frac{\mathcal J_2}{f_2^2}\right)^2+\frac{\partial_z\rho^2}{\rho^2}{z'}^2&=0~.
\end{align}

We specialize to geodesics which are stationary on $\Sigma$, such that $z(\lambda)=z_\star$. This leads to
\begin{align}
	\frac{f_4^2}{f_1^2}j_1^2+\frac{f_4^2}{f_2^2}j_2^2&=1~,
	&
	j_i&\equiv\frac{\mathcal J_i}{\mathcal E}~,
	\label{eq:null3}
	\\
	\frac{\partial_{\bar{z}}f_4^2}{f_4^4}-\frac{\partial_{\bar{z}}f_1^2}{f_1^4}j_1^2 -\frac{\partial_{\bar{z}}f_2^2}{f_2^4}j_2^2&=0~,
\end{align}
all evaluated at $z_\star$.
These are one real and one complex equation. They can be used to determine the real $j_i^2$, leaving a constraint on $z_\star$ which singles out curves in $\Sigma$ on which $z_\star$ can lie.
We find
\begin{align}\label{eq:geod-eq-gen}
	j_1^2&=\frac{f_1^4}{f_4^4}\frac{\partial(f_4^2/f_2^2)}{\partial(f_1^2/f_2^2)}~,
	&
	j_2^2&=\frac{f_2^4}{f_4^4}\frac{\partial(f_4^2/f_1^2)}{\partial(f_2^2/f_1^2)}~,
	&
	\Im\left[\frac{\partial(f_4^2/f_2^2)}{\partial(f_4^2/f_1^2)}\right]&=0~.
\end{align}
The last equation constrains $z_\star$. The general null geodesics fixed on $\Sigma$ are obtained by solving the last equation, which guarantees that $j_i^2$ are real.\footnote{This can be seen by rewriting the denominators in $j_i^2$ using $\partial(f_1^2/f_2^2)=f_1^2/f_4^2 \partial( f_4^2/f_2^2)+f_4^2/f_2^2\partial(f_1^2/f_4^2)$.}
A further constraint is that $j_1^2$ and $j_2^2$ have to be non-negative.
We note that
\begin{align}\label{eq:f12df4}
	\frac{f_4^2}{f_1^2}&=\frac{N_1}{h_1^2|W|}=1+\frac{2h_2|\partial h_1|^2}{h_1|W|}~, &
	\frac{f_4^2}{f_2^2}&=\frac{N_2}{h_2^2|W|}=1+\frac{2h_1|\partial h_2|^2}{h_2|W|}~,
\end{align}
where we used that $W$ is negative for regular solutions, so $|W|=-W$. 
This implies $f_4^2/f_i^2\geq 1$, which with (\ref{eq:null3}) implies $j_i^2\leq 1$ and $j_1^2+j_2^2\leq 1$.

The condition on $z_\star$ in (\ref{eq:geod-eq-gen}) can be written as
\begin{align}
	0&=\Im\left[\partial(f_4^2/f_2^2)\bar\partial(f_4^2/f_1^2)\right]
	=\Im\left[\partial\left(\frac{2h_1|\partial h_2|^2}{h_2W}\right)\bar\partial\left(\frac{2h_2|\partial h_1|^2}{h_1W}\right)\right],
\end{align}
which is equivalent to 
\begin{align}
	0&=\Im\left[\partial\left(\frac{h_2W}{h_1|\partial h_2|^2}\right)\bar\partial\left(\frac{h_1W}{h_2|\partial h_1|^2}\right)\right]
\end{align}
Using the definition of $W$ the condition becomes
\begin{align}\label{eq:geod-F}
	0&=\partial\left(F+\bar F\right)\bar\partial\left(\frac{1}{F}+\frac{1}{\bar F}\right)-{\rm c.c.}
	=\left(\frac{1}{F^2}-\frac{1}{\bar F^2}\right)\left(\partial F \bar\partial\bar F-\bar\partial F\partial\bar F\right), & F&=\frac{h_2\partial h_1}{h_1\partial h_2}~.
\end{align}
We find two types of solutions: Demanding the first factor to vanish gives a condition of first order in derivatives of $h_{1/2}$. These are geodesics satisfying a BPS condition. Demanding the second factor to vanish leads to (\ref{eq:geod-gen}) for more generic geodesics.

\bigskip
\textbf{BPS geodesics:} The condition for BPS geodesics resulting from (\ref{eq:geod-F}), $F=\pm \bar F$, upon taking into account regularity constraints, leads to
\begin{equation}
	\partial h_1\bar{\partial}h_2-\bar{\partial}h_1\partial h_2=0. \label{eq:cond}
\end{equation}
This shows \eqref{eq:h1h2-cond-gen} in the main part and amounts to $\cY$ in (\ref{eq:b1b2}) vanishing.
This one real condition constrains the geodesics to be located at a point along a curve through $\Sigma$. It implies
\begin{align}\label{eq:bps-null-geod}
	f_1^2+f_2^2-f_4^2&=0~, & \partial_z(f_1^2+f_2^2-f_4^2)&=0~.
\end{align}
This leads to the angular momentum fractions in (\ref{eq:geod-eq-gen}) given by
\begin{align}
	\frac{\mathcal J_i}{\mathcal E}=j_i&=\pm\frac{f_i^2}{f_4^2}~,
	&
	|j_1|+|j_2|&=1~.
\end{align}
Together these relations imply that the null and geodesic equations are satisfied.
The last equation is the BPS condition.

Special cases arise when the curve singled out by the condition \eqref{eq:cond} approaches a boundary of $\Sigma$. On each boundary one of $h_{1/2}$ vanishes while the other satisfies a Neumann boundary condition. Locally the boundary can be taken as a segment of the real line, $z=x+iy$ with the boundary at $y=0$. Expanding \eqref{eq:cond} near $y=0$ then leads to the conditions in (\ref{eq:nice-cond-1}), (\ref{eq:nice-cond-2}).
The algebraic relation in (\ref{eq:bps-null-geod}) still holds with one $f_i^2$ vanishing. This extends to the derivative relation in (\ref{eq:bps-null-geod}): the vanishing warp factor behaves as $f_i^2\sim y^2\rho^2$ for the corresponding $S^2$ to close off without conical singularity at $y=0$, so the corresponding $\partial_z f_i^2$ vanishes as well.

\bigskip
\textbf{Penrose limit on $\partial\Sigma$:} For the Penrose limit we start with geodesics on the boundary of $\Sigma$, where one of the harmonic functions vanishes. 
We choose the boundary where $h_1$ vanishes.
With the coordinate transformation (\ref{eq:Penrose-dSigma-coord}) the metric in (\ref{eq:met1}) becomes
\begin{align}
	ds^2=\,&-4dx^+ dx^-+dr^2+r^2ds^2_{S^2} +dx_8^2+x_8^2ds^2_{S_1^2}+dx_6^2+dx_7^2+dx_8^2
	\nonumber\\ &
	-(dx^+)^2
	\left(r^2+x_6^2\right)
	-\frac{(dx^+)^2}{8\rho^2}
	\left(
	(\zeta^2 \partial_z^2+\zeta\bar\zeta\partial_z\partial_{\bar z})(f_2^2-f_4^2)+\rm{c.c.}
	\right),
\end{align}
where $\zeta=x_7+ix_8$.
The derivatives can be evaluated noting that using (\ref{eq:bps-null-geod}) we have $\partial_z^2(f_2^2-f_4^2)=\partial_z^2[f_4^2(f_2^2/f_4^2-1)]=f_4^2\partial_z^2(f_2^2/f_4^2-1)$. The expressions in (\ref{eq:f12df4}) can then be used. This leads to the metric in (\ref{eq:ppwave-metric-dSigma}).

The evaluation of the dilaton leading to (\ref{eq:Penrose-dSigma-dilaton}) is straightforward. For the 3-form field strengths we find, using $\vol_{S^2_i}=\cos\theta_i d\theta_id\phi_i$,
\begin{align}
	H_{(3)}&=\vol_{S_1^2}\wedge\left[\frac{d\zeta \partial_z b_1+d\bar\zeta\partial_{\bar z}b_1}{2\rho}+\frac{d\zeta \zeta\partial_z^2b_1+(d\zeta\bar\zeta+d\bar\zeta \zeta)\partial_z\partial_{\bar z}b_1  +d\bar\zeta\bar\zeta\partial_{\bar z}^2b_1}{4\rho^2}\right],
	\nonumber\\
	F_{(3)}&=dx_6\wedge dx^+\wedge \frac{d\zeta\partial_z b_2+d\bar\zeta\partial_{\bar z}b_2}{2 f_2\rho}\,.
\end{align}
For the boundary where $h_2$ vanishes the roles of $H_{(3)}$ and $F_{(3)}$ are exchanged.
Evaluating them using the boundary conditions and the conditions satisfied by $z_\star$ shows that both vanish in the Penrose limit.
For the 5-form field strength we find, with $\vol_{AdS_4}=\cosh\hat\rho\sinh^2\!\hat\rho\,d\hat\rho\wedge dt\wedge \vol_{S^2}$,
\begin{align}
	F_{(5)}&=w+\star w~, 
	&
	w&=-2r^2dr\wedge dx^+\wedge \vol_{S^2}\frac{d\zeta\partial_z j_1+d\bar\zeta\partial_{\bar z}j_1}{f_4^3\rho}\,.
\end{align}
Evaluating this explicitly leads to $\partial_z j_1/(f_4^3\rho)\vert_{z=z_\star}=-1$.

\bigskip
\textbf{Penrose limit in ${\rm int}(\Sigma)$:} We now turn to the generic case of geodesics in the interior of $\Sigma$ where $z_\star$ satisfies (\ref{eq:h1h2-cond-gen}). The pp-wave geometry can be obtained from the coordinate transformation
\begin{align}\label{eq:Penrose-gen-transf}
	t&=x^+ +\frac{x^-}{f_4^2}~, &
	\phi_1&=x^+ -\frac{x^-}{f_4^2} +\frac{f_2}{f_1f_4} x^9~,&
	\phi_2&=x^+ -\frac{x^-}{f_4^2}- \frac{f_1}{f_2f_4} x^9~,
	\nonumber\\
	\hat\rho&=\frac{r}{f_4}\,,
	&
	\theta_1&=\frac{x_4}{f_1}\,,
	&
	\theta_2&=\frac{x_6}{f_2}\,,
	&
	z&=z_\star+\frac{\zeta}{2\rho}\,.
\end{align}
We assume that the metric functions are homogeneously large, implement this coordinate transformation in the metric (\ref{eq:met1}) and drop terms which scale as inverse powers of the metric functions.
This leads to the pp-wave metric
\begin{align}\label{eq:Penrose-gen-ds2}
	ds^2=\,&-4dx^+ dx^-+dr^2+r^2ds^2_{S^2}+dx_4^2+dx_6^2+|d\zeta|^2+dx_9^2
	-(dx^+)^2(r^2+x_4^2+x_6^2)
	\nonumber\\
	&+\frac{dx^+ dx^9}{f_1f_2f_4\rho}\left[\zeta(f_2^2\partial f_1^2-f_1^2\partial f_2^2)+\rm{c.c.}\right]
	+\frac{(dx^+)^2}{8\rho^2}\left[
	(\zeta^2\partial^2+\zeta\bar\zeta\partial\bar\partial)(f_1^2+f_2^2-f_4^2)+\rm{c.c.}
	\right]~,
\end{align}
where all metric functions and derivatives are evaluated at $z_\star$.
The derivatives of the metric functions evaluated at $z_\star$ can be rewritten using (\ref{eq:bps-null-geod}) as follows,
\begin{align}
	\partial^2(f_1^2+f_2^2-f_4^2)&=f_4^2\partial^2\left(\frac{f_1^2}{f_4^2}+\frac{f_2^2}{f_4^2}-1\right)\,,
	&
	f_2^2\partial^2f_1^2-f_1^2\partial^2f_2^2&=f_2^4\partial\left(\frac{f_1^2}{f_2^2}\right)\,.
\end{align}
One can then use (\ref{eq:f12df4}) to evaluate the expressions.
The dilaton in (\ref{eq:met1}) becomes
\begin{align}
	e^{4\phi}&=-\frac{h_2\partial h_2}{h_1\partial h_1}~.
\end{align}
This is real thanks to \eqref{eq:cond}.
For the 3-form field strengths (\ref{eq:H3F3F5}) we find, with $\vol_{S^2_i}=\cos\theta_i d\theta_id\phi_i$,
\begin{align}
	H_{(3)}&=dx^4\wedge dx^+\wedge\frac{(\partial_z b_1)_{z=z_\star}d\zeta+\rm{c.c.}}{2f_1\rho}
	~,
&
	F_{(3)}&=dx^6\wedge dx^+\wedge\frac{(\partial_z b_2)_{z=z_\star}d\zeta+\rm{c.c.}}{2f_2\rho}~.
\end{align}
We note that homogeneously large metric functions need $h_{1/2}$ large, and when $h_{1/2}\rightarrow \alpha h_{1/2}$ the metric functions scale as $f_i^2\rightarrow \alpha f_i^2$. Since the $b_i$ scale like $h_{1/2}$, $b_i/(f_i\rho)$ is order one.
For the 5-form field strength in (\ref{eq:H3F3F5}) we find, with $\vol_{AdS_4}=\cosh\hat\rho\sinh^2\!\hat\rho\,d\hat\rho\wedge dt\wedge \vol_{S^2}$,
\begin{align}
	F_{(5)}&=\omega+\star\omega~, & \omega&=-2dr\wedge dx^+\wedge \vol_{S^2}\wedge\frac{(\partial_z j_1)_{z=z_\star}+\mathrm{c.c.}}{f_4^3\rho}~.
\end{align}
Since $j_1$ scales like $h_{1/2}^2$, this is an order one expression.
Noting that $\cY\vert_{z=z_\star}=0$ we have
\begin{align}
	(\partial_z b_1)_{z=z_\star}&=\frac{2h_1^2h_2}{N_1}\partial_z\cY+2\partial_z h_2^D\Big\vert_{z=z_\star}\,,
	&
	(\partial_z b_2)_{z=z_\star}&=\frac{2h_1h_2^2}{N_2}\partial_z\cY-2\partial_z h_1^D\Big\vert_{z=z_\star}\,,
	\nonumber\\
	(\partial_z j_1)_{z=z_\star}&=3\partial_z\left(\cC-\cD\right)+\frac{h_1h_2}{W}\partial_z\cY\Big\vert_{z=z_\star}\,.
\end{align}

\bibliography{Penrose}
\end{document}